\documentclass[%
 reprint,
 amsmath,amssymb,
 jap,
]{revtex4-1}

\usepackage{graphicx}
\usepackage{dcolumn}
\usepackage{bm}
\usepackage{xcolor}
\usepackage{siunitx}

\newcommand{\beginsupplement}{%
        \setcounter{table}{0}
        \renewcommand{\thetable}{S\arabic{table}}%
        \setcounter{figure}{0}
        \renewcommand{\thefigure}{S\arabic{figure}}%
     }

\begin{document}


\title{Universal Characterisation of Cavity--Magnon Polariton Coupling Strength Verified in Modifiable Microwave Cavity}

\author{Jeremy Bourhill}
\email{jeremy.bourhill@imt-atlantique.fr}
\author{Vincent Castel}
 \affiliation{IMT Atlantique, Technopole Brest-Iroise, CS 83818, 29238 Brest Cedex 3, France}
 \affiliation{%
Lab-STICC (UMR 6285), CNRS, Technopole Brest-Iroise, CS 83818, 29238 Brest Cedex 3, France}%


\author{Alexandre Manchec}
\author{Gwendal Cochet}
\affiliation{Elliptika (GTID), 29200 Brest, France}%
\homepage{http://www.elliptika.com/}

%

\date{\today}

\begin{abstract}
A comprehensive study of the frequency dependence of the photon-magnon coupling for different magnetic samples is made possible with a tuneable 3D--printed re--entrant cavity. Strong coupling is achieved, with values ranging between 20--140 MHz. The reworked theory, experimentally verified for the first time here, enables coupling values to be calculated from simulations alone, enabling future experiments with exotic cavity designs to be precisely engineered, with no limitations on sample and cavity geometry. Finally, the requirements of the deep strong coupling regime are shown to be achievable in such experiments.

\end{abstract}

\maketitle


\section{\label{sec:1}Introduction}

The past decade has seen rapid development in the field of cavity magnonics. As far as light--matter interactions go, magnetic materials are attractive given their exceptionally large density of spins and low losses. The former makes them easy to couple light to, whilst the latter gives long coherence times. In addition, the  collective spin wave resonance (quantised as the magnon) is widely tuneable with an external magnetic field, therefore interactions (i.e. with a qubit) can be generated for any desired frequency. The strong coupling regime, in which the magnon-photon coupling rate ($g$) is greater than the cavity and magnon loss rates is easily achievable in these systems even at room temperature. This regime is a minimum requirement for the coherent transfer of quantum information. For these reasons, a great deal of research has aimed at developing these systems with the application of quantum information processing and memory in mind \cite{Chumak14,Chumak15,Tanji09,Zhang15,wesenberg}. A hybrid system combining the fast manipulation rates of superconducting qubits with the assets of magnonic systems, through strong coupling with a microwave cavity, presents as a promising solution for such applications \cite{Tabuchi405,Tabuchi16,Nakamura_Review}. 

In addition, due to the possibility of coupling magnon modes to photons at optical frequencies \cite{Nakamura_Review,Zhang2,Osada,Shen,Demokritov}, magnon systems may be considered as a candidate for coherent conversion of microwave and optical photons. In addition, magnons interact with elastic waves \cite{Nakamura_Review,Zhang3} permitting the combination of mechanical and magnetic systems. These systems therefore possesses great potential as an information transducer that mediates interconversion between information carriers of different physical nature.

On a more fundamental level, cavity--magnon systems have allowed the study of strong coupling physics \cite{Liberato2,Rameshti2}, with a handful of experiments demonstrating the ultrastrong coupling (USC) regime \cite{Flower19,Bourhill16,goryachev14,Zhang1}, in which the coupling rate is an appreciable proportion of the system frequency (i.e. $g/\omega>0.1$). New and interesting physics has been discovered in these regimes \cite{Liberato2,Auer,Liberato3,Liberato5,Lambert1,Nataf1,ciuti1}, in which the rotating wave approximation breaks down. One regime still relatively unexplored experimentally however, is the deep strong coupling (DSC) regime, in which the coupling energy is larger than the system energy, i.e. $g>\omega$. In this regime non-classical states can be generated such as Schr\"odinger-cat and squeezed states \cite{DSC,Barberena:2017aa}.

Parallel to cavity magnonics, but now converging, is the field of spintronics, in which an electrical current pumping (readout) of spin waves is achieved through the (inverse) spin Hall effect. This effect relies on ferromagnet--normal metal heterostructures and has been demonstrated for a wide variety of material combinations in thin film form \cite{Kajiwara2010,Wang1,czeschka,castel2}. There now exists the real possibility of combining the two fields to enable the transmission and electrical read out of quantum states in ferromagnets using a hybrid architecture \cite{Maier-Flaig1,Bai1,Cao1}. Given the necessary thin film nature however, it is difficult to achieve uniform RF fields over the sample volume that will result in large photon-magnon coupling values in typical cavities. This non-uniformity of field causes issues of coupling to unwanted, non-uniform, higher order magnon modes \cite{Bourhill16,Zhang4}, and non-applicability of the commonly used theoretical models \cite{Tabuchi113,Zhang1}.

It would be ideal to have some predictive capacity for exotic cavity designs in which the RF field is not necessarily uniform over the magnetic sample in order to enhance coupling values through RF magnetic field focusing \cite{goryachev14}. Majority of treatments found within the literature operate under  specific assumptions (such as field uniformity or spherical sample geometry). Such a predictive capacity would permit one to design experiments with full knowledge of cavity mode frequencies and coupling strengths, regardless of cavity and sample geometry. Thus, hybrid mode frequencies can be engineered to coincide with other resonances within the system, whether they be additional cavity modes or modes of a different physical nature, which will greatly enhance the efficiencies of spin transfer protocols between distinct magnetic samples \cite{Zhang15,Rameshti,PhysRevLett.118.217201,PhysRevB.100.094415,PhysRevA.93.021803} and information interconversion \cite{Nakamura_Review}. 


Such a model has been previously developed \cite{Flower19} but failed to accurately predict $g$ values for thin film samples. A capable predictive model has also been developed for paramagnetic systems \cite{weichselbaumer}. Here, using a large set of coupling values to thin film magnetic samples inside a modifiable cavity, as well as results taken from the literature, we rigorously demonstrate the presented model{'}s capability to be applied universally to any cavity--magnon system. The ability to compare different experiments is only made possible by the general nature of this model, compared to others that are only applicable for a particular system or class of systems \cite{Zhang1,Zhang4,Liu,Rameshti,Cao1,Maier-Flaig1,Boventer,Tabuchi113}. Finally, using this analysis, we demonstrate the requirements of cavity magnon experiments to reach the DSC regime. 


\section{\label{sec:2}Theory}

Cavity magnon polaritons are bosonic quasiparticles associated with the hybridisation of a photon and a magnon within a cavity resonator. In the strong coupling regime, this results in an anti--crossing in the dispersion spectrum as the magnon frequency is tuned close to the cavity frequency by application of an external DC magnetic field, $H_0$. The Hamiltonian for a magnon/cavity photon system (not in the DSC regime) can be described by the Tavis-Cummings model \cite{TC} and can be found elsewhere \cite{TC,goryachev14,Fink,Huebl,Soykal}. The Hamiltonian describes two coupled harmonic oscillators, where the eigenfrequencies of the split mode are: 
	\begin{equation}
		\omega_\pm=\sqrt{\frac{\omega_c^2+\omega_m^2}{2}\pm\sqrt{\left(\frac{\omega_c^2-\omega_m^2}{2}\right)^2+4\omega_c\omega_mg_{cm}^2}},
		\label{eq:pm}
	\end{equation}
where $\omega_c$ is the cavity frequency in angular units, $g_{cm}$ the coupling strength as a measure for the frequency of coherent information exchange, which determines the size of the mode splitting, and $\omega_m^\text{sphere}(H_0)=\gamma H_0,~\text{and}~\omega_m^\text{film}(H_0)=\gamma\sqrt{H_0(H_0+M_s)}$ are the magnon resonant frequencies as a function of $H_0$ and material dependent constants; saturation magnetisation $M_s$ and gyromagnetic ratio $\gamma$. Here $\omega_m^\text{film}$ is for in-plane magnetisation. 

A succinct derivation of the formula for the coupling rate $g_{cm}$ is given by Flower \textit{et al.} \cite{Flower19}, and appears as:
	\begin{equation}
		\frac{g_{cm}}{2\pi}=\frac{\gamma}{4\pi}\eta\sqrt{\frac{\mu_0S\hbar\omega_c}{V_m}}=\eta\sqrt{\omega_c}\frac{\gamma}{4\pi}\sqrt{\frac{\mu}{\text{g}\mu_B}\mu_0\hbar n_s},
		\label{eq:g}
	\end{equation}
where $S=\frac{\mu}{\text{g}\mu_B}N_s$ is the total spin number of the macrospin operator, $\mu_B$ is the Bohr magneton, $\mu=5\mu_B$ \cite{Gilleo,Xie} is the magnetic moment for YIG, $\text{g}=2$ is the Land\'e g-factor for an electron spin, $N_s$ the number of spins in the sample, $V_m$ the magnetic sample volume, $\mu_0$ is the vacuum permeability, $\hbar$ the reduced Planck{'}s constant and $n_s=N_s/V_m$ the density of spins. 

It is worthwhile clarifying the apparent confusion surrounding the value of $n_s$ for YIG. To the authors{'} knowledge, there appears three different values of $n_s$ quoted throughout the literature: $4.22\times10^{27}$ m$^{-3}$ \cite{Zhang1,Zhang4,Liu}, $2.11\times10^{28}$ m$^{-3}$ \cite{Flower19,goryachev14,Rameshti,Cao1} and $2.11\times10^{28}\mu_B$ m$^{-3}$ \cite{Maier-Flaig1,Boventer,Tabuchi113}, and despite the appearance of $\mu_B$ in the latter value, some authors appear to simply neglect it when using their equivalent eq.(\ref{eq:g}) as if they were using the second quoted value. The correct value to use to calculate $g_{cm}$ from eq.(\ref{eq:g}) is $n_s=4.22\times10^{27}$ m$^{-3}$. This can be derived from $n_s=(\frac{1}{8}a^3)^{-1}$ where $a=12.376$ \si{\angstrom} is the lattice constant of the cubic unit cell and the $\frac{1}{8}$ transforms to the primary cell \cite{Xie,spinwavebook}. Note that multiplying by $5\mu_B$ would net a result of $2.11\times10^{28}\mu_B$ cm$^{-3}$, but is not necessary as it is already taken into account in eq.(\ref{eq:g}) by the appearance of $\mu$ in the $S$ term. 

The so-called {``}form--factor{''}, $\eta$ describes the proportion of the cavity mode{'}s magnetic field perpendicular to $H_0$ ($\vec{H}.\vec{e_x}$ and $\vec{H}.\vec{e_y}$, assuming $H_0$ is applied along the $z$-axis) as well as the proportion of this perpendicular field located within $V_m$ compared to the entire cavity volume, $V_c$ \cite{Flower19}. It can be defined as:
	\begin{equation}
		\eta=\sqrt{\frac{\left(\int_{V_m} \vec{H}.\vec{e_x} dV\right)^2+\left(\int_{V_m} \vec{H}.\vec{e_y} dV\right)^2}{V_m\int_{V_c} |H|^2 dV}}.
		\label{eq:eta}
	\end{equation}
It can be seen from the form of eq.(\ref{eq:eta}) that the case of $\eta=1$ would correspond to all of the applied RF cavity magnetic field being perpendicular to $H_0$ as well as being totally confined to $V_m$. The latter condition represents a divergence from the definition of $\eta$ given in previous publications \cite{Zhang1,Tabuchi113,Zhang15} and in doing so extends the applicability of eq.(\ref{eq:g}) to systems in which RF fields are non-uniform over $V_m$. 

As highlighted in \cite{Flower19,Huebl}, $g_{cm}$ only depends on $n_s$, $\mu$, $\eta$ and $\omega_c$, and therefore an increase in $N_s$ does not necessarily facilitate an increase in $g_{cm}$. Take for example the case of a large cavity with localised field in a small volume $V_f<V_c$, and assume $V_f=V_m$. An increase in $V_m$ and hence $N_s$ will in fact reduce $g_{cm}$ as the ratio of the integrals in eq.(\ref{eq:eta}) remains unchanged but $V_m$ has increased. 

The form of equation (\ref{eq:eta}) is derived from first principles \cite{Flower19} and its use of sample volume $V_m$ as opposed to cavity or {``}mode{''} volume allows the simplification of equation (\ref{eq:g}) such that there is only dependence on spin density $n_s$. This permits the direct comparison of all cavity magnon experiments regardless of the size or type of cavity used, and allows equation (\ref{eq:g}) to be solved for any cavity geometry. This new form of $\eta$ presented in equation (\ref{eq:eta}) is therefore a more robust and universal form for the characterisation and prediction of cavity-magnon systems.


\section{\label{sec:3}Experiment Description}
\begin{figure}[t!]
		\centering
		\includegraphics[width=0.5\textwidth]{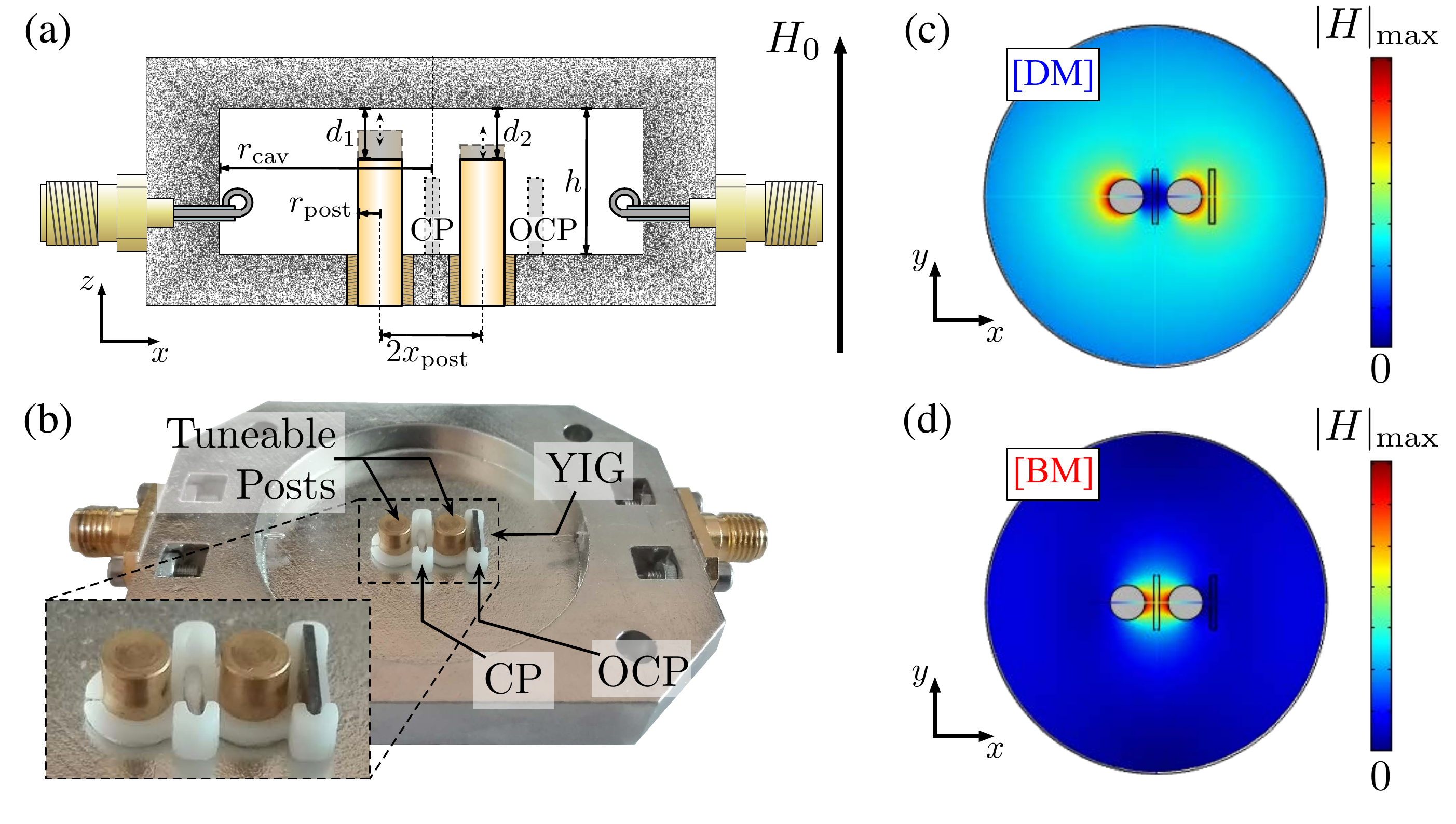}
		\caption{(a) Illustration of the double post re-entrant cavity used (b) Photograph of the cavity, with the two positions of the magnetic sample within the 3D printed sample holder (not metallised) demonstrated. (c) and (d): Magnetic field intensity for the DM and BM showing field concentration at the OCP and CP thin film positions, respectively.}
		\label{fig:1}
	\end{figure}
\begin{figure}[h!]
		\centering
		\includegraphics[width=0.5\textwidth]{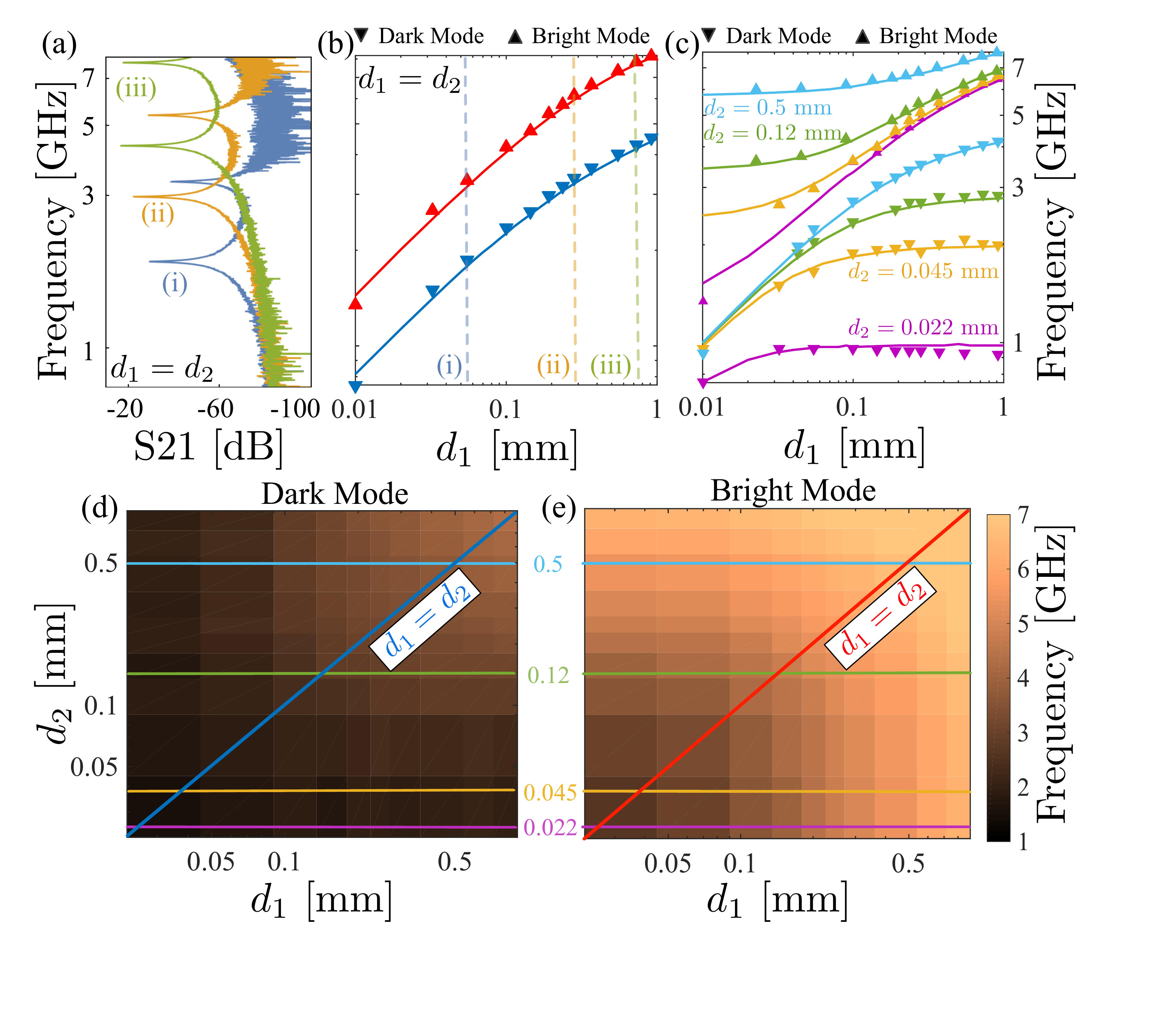}
		\caption{(a) S21 spectra for different $d_1=d_2$, the values of which are indicated by the vertical dashed lines (i) to (iii) in (b). (b) Frequency tuning of the DM and BM as functions of $d_1$, when $d_1=d_2$, and (c) when $d_1\neq d_2$ for given $d_2$ values, with both experimental and simulated results plotted. (d) and (e) Density plots of the experimentally measured resonant frequency of the DM and BM, respectively as a function of $d_1$ and $d_2$ with the diagonal lines of $d_1=d_2$ representing the {``}paths{''} traversed in (b) and the horizontal lines of constant $d_2$ representing those traversed in (c). }
		\label{fig:2}
\end{figure}
The cavity used (Fig.~\ref{fig:1}(a)-(b)) is a tuneable realisation of a double post re-entrant cavity, which extracts greater coupling values for a given sample than could be achieved from a standard waveguide due to a focusing of magnetic field. These cavities, an extension of the single--post form \cite{Fujisawa,Eshbach,LeFloch}, were initially developed by Goryachev \textit{et al.} \cite{goryachev14} and have since become more commonplace throughout the literature \cite{Boventer,Flower19,goryachev18,castel19,Clai}. 

The cavity was 3D-printed and metallised by Elliptika$\textsuperscript{TM}$ using the technique described in \cite{castel19}. The two posts are commercially purchased and inserted subsequent to printing, as is the thread in the cavity base into which the posts are screwed. There are two first-order resonant modes of such a cavity, termed the dark and bright modes (DM and BM: Fig.~\ref{fig:1}(c)--(d)). Both contain the mode{'}s electric field between the top of the post and the roof of the cavity. For the DM, the $E$-fields are in-phase, resulting in the circulating $H$-fields destructively interfering in the region between the posts (hence {``}dark{''}), whilst the opposite is true for the BM \cite{goryachev14}. The distances between the posts and the cavity roof, $d_1$ and $d_2$ (as depicted in Fig.~\ref{fig:1}(a)), therefore define the capacitance of the cavity and hence determine the resonant frequencies of the modes. 

We use two magnetic samples of YIG; a thin film sample with in-plane magnetisation and a sphere. For the film sample,  $9~\mu$m of YIG has been grown by liquid phase epitaxy on both sides of a GGG substrate ($6.5\times3.8\times0.5$) mm$^3$. The sphere has a radius of $235~\mu$m. They can be placed in either the central position (CP) of the cavity, between the two posts or in the off-centre position (OCP).
The thin film sample is positioned using a 3D printed sample holder, seen in Fig.~\ref{fig:1}(b), placed around the tuneable posts. 

The cavity has radius $r_\text{cav}=20$~mm and height $h=4.6$~mm, whilst the posts have radius $r_\text{post}=2.05$~mm and are located at $x_\text{post}=3.4$~mm from the cavity centre. These parameters were chosen, given the commercially purchased post dimensions, to maximise geometric factor for a given operating frequency, whilst still maintaining a broad frequency tuning range. 

The full tuning dynamics of the cavity are plotted in Fig.~\ref{fig:2}, which illustrates the exceptional range made possible by screwing the posts into or out of the cavity. The transmission spectra of the cavity for different $d_1=d_2$ values is shown in Fig.~\ref{fig:2}(a) demonstrating an ideal, clean spectra for filtering applications, whilst the agreement between simulation and experiment for $d_1=d_2$ tuning up to $\sim1$~mm is shown in Fig.~\ref{fig:2}(b). It can be seen in Fig.~\ref{fig:2}(c)--(e) that depending on the respective values of $\{d_1,d_2\}$, there is a large array of DM and BM resonant frequency pairings available. The versatility and predictability with which both mode frequencies can be positioned is a large advantage of this cavity design for numerous applications in which precise locations of frequencies might be important. $Q$-factors of the modes depend on the post heights (see Supp. Mat.) and range in values from $150-800$.


The cavity is placed inside a DC magnet with the field, $H_0$ parallel to the posts of the cavity and therefore in the plane of the thin film sample (Fig.~\ref{fig:1}(a)-(b)). The magnetic field is swept such that the magnon resonance passes over both modes (by viewing the scattering parameters of the cavity on a Vector Network Analyser), then the mode frequencies are changed by tuning the post heights equally and the field swept again, and so on. In this way a multi-frequency readout of magnon-photon couplings can be achieved using a single cavity, without need for disassembly or re-calibration of instruments.

\section{\label{sec:4}Results}
\begin{figure}[h!]
		\centering
		\includegraphics[width=0.5\textwidth]{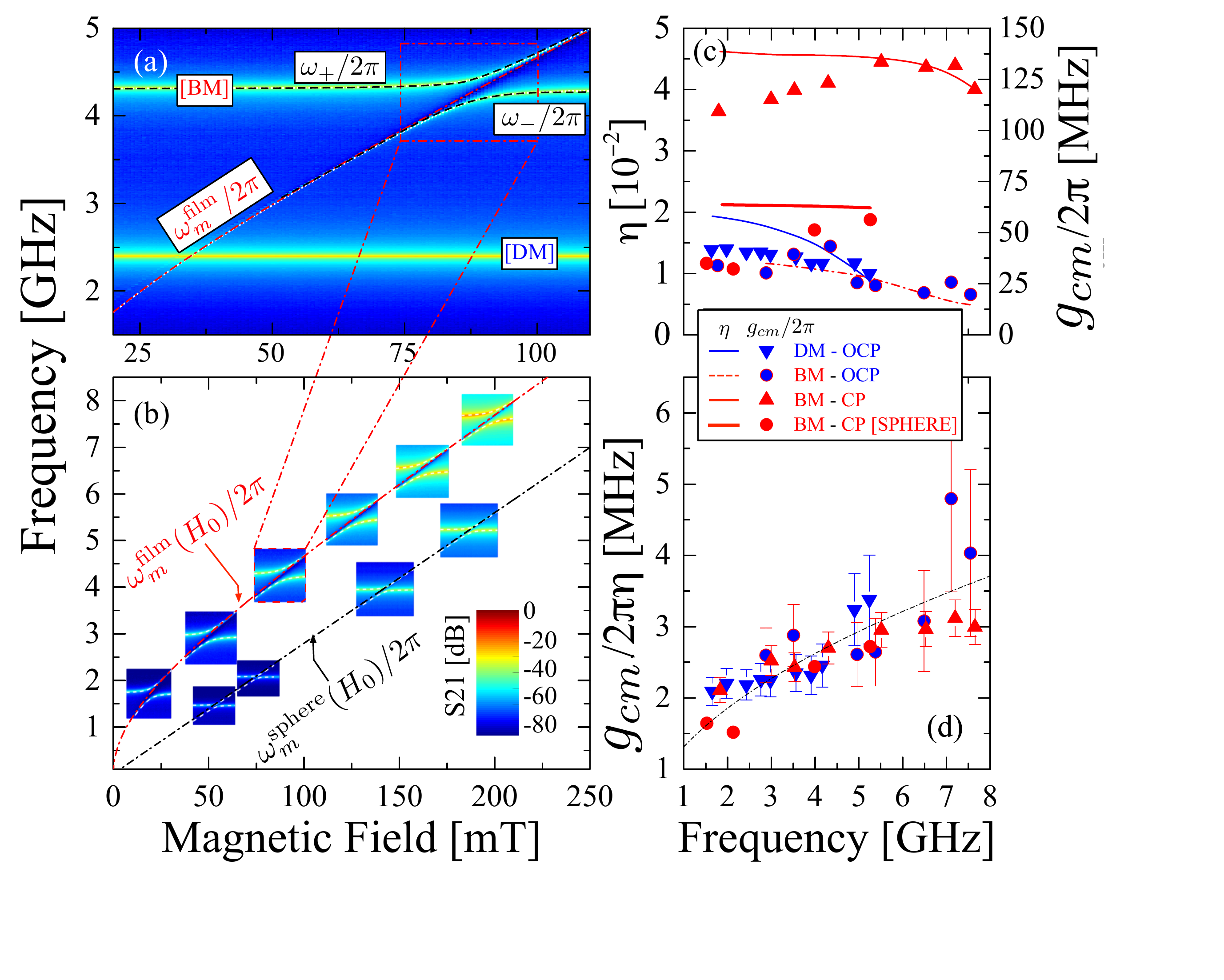}
		\caption{(a) S21 spectra of the cavity at a fixed $\{d_1,d_2\}$ setting for the thin film sample located in the CP. Plotted in red is the $\omega_m(H_0)/2\pi$ curve for a thin film sample. (b) A collection of CP thin film and sphere coupling to the BM S21 spectra as a function of magnetic field. Each spectra is a separate experimental run with a unique $\{d_1,d_2\}$ setting. (c) $g_{cm}/2\pi$ as a function of mode frequency extracted from the spectra in (b). (d) $g_{cm}/2\pi\eta$ as a function of mode frequency with the fitted line given by eq.(\ref{eq:g}).}
		\label{fig:3}
	\end{figure}
The position of the centre of each anti--crossing can be used to determine $\gamma=2\pi\times 28$ GHz/T and $M_s=0.1775$ T. The magnon linewidth can be measured when it is tuned half-way between the DM and BM; $\sim2.5~$MHz and $\sim3.5$ MHz for the thin film and sphere samples, respectively. Then, values of $g_{cm}$ can be extracted from each of the individual density plots for a given $\{d_1,d_2\}$ (an example given in Fig.~\ref{fig:3}(a)) by fitting eq.(\ref{eq:pm}). Finally a comprehensive survey of coupling as a function of frequency and $\eta$ for both samples can be conducted, the results of which are presented in Fig.~\ref{fig:3}(b)--(d). 

The form factor $\eta$ is different depending on whether the DM or BM are used, as well as the values of $\{d_1,d_2\}$ and the sample geometry. The dependence of $\eta$ on gap size is calculated using Finite Element Modelling (FEM) software COMSOL$\textsuperscript{TM}$ for each of the four unique cases: DM and BM for the thin film in the OCP, BM for the thin film in the CP (coupling with DM was too small to measure) and BM for the sphere in the CP (see Supp. Mat.). Specific $\eta$ values for each individual $\{d_1,d_2\}$ value can then be interpolated from the four resulting functions. 
	
If one normalises the $g_{cm}$ values and both sides of eq.(\ref{eq:g}) by $\eta$ we observe agreement between experiment and theory, as shown in Fig.~\ref{fig:3}(d) for all samples and both positions. Experimental errors here are due to mesh and dimension errors in the FEM process used to determine $\eta$.

\section{\label{sec:5}Discussion}
\begin{figure}[t!]
		\centering
		\includegraphics[width=0.5\textwidth]{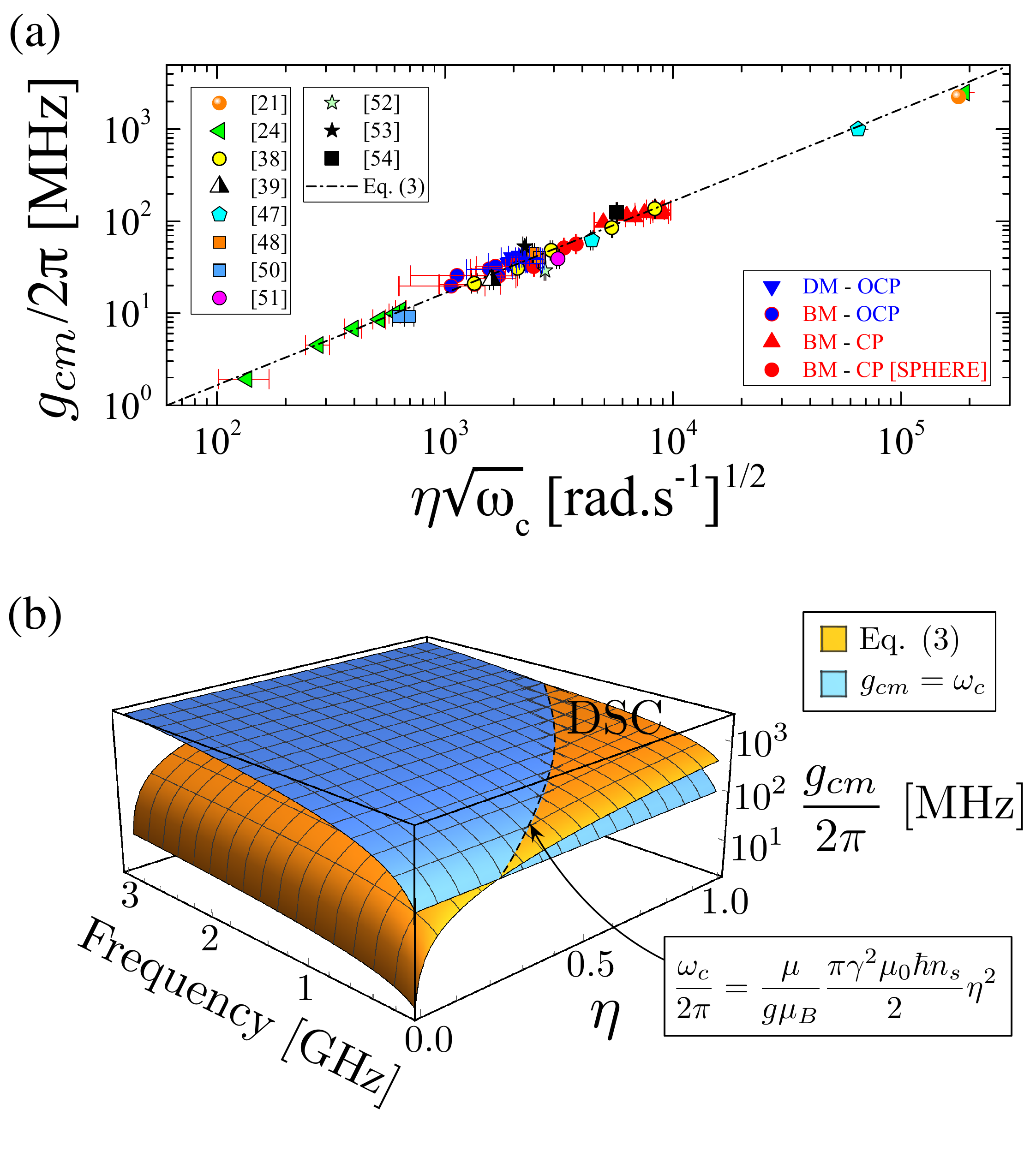}
		\caption{(a) Cavity--magnon coupling strength as a function of $\eta\sqrt{\omega_c}$ for the thin film and sphere experimental results presented in this paper as well as a host of other experimental results taken from the literature \cite{Flower19,Zhang1,Zhang4,Tabuchi113,goryachev18,castel19,NatCom8,Harder,Boventer,PhysRevB.99.134445,goryachev2017strong}. (b) Eq.(\ref{eq:g}) and $g_{cm}=\omega_c$ plotted as a function of $\omega_c/2\pi$ and $\eta$, revealing the parameter requirements for DSC in which $g_{cm}>\omega_c$.}
		\label{fig:4}
	\end{figure}

Compared to majority of the literature{'}s expressions for $g_{cm}$, the uniqueness of eq.(\ref{eq:g}) is that there is no explicit dependence on the number of spins, $N_s$, as we work in density, $n_s$. Nor is there dependence on the volume of the cavity or {``}mode volume{''}. Most of this information is handled and or cancelled out by our form of $\eta$. This means that the only experimentally dependent variables in eq.(\ref{eq:g}) are $\eta$ and $\omega_c$, assuming consistent material constants. Therefore by plotting $g_{cm}/2\pi$ against $\eta\sqrt{\omega_c}$ for published cavity-magnon experiments using YIG, all such experiments can be directly compared to those derived here, as demonstrated in Fig.~\ref{fig:4}(a).

Included in Fig.~\ref{fig:4}(a) are both spherical and thin film geometries, where the $\eta$ values have been calculated using FEM from the quoted sample and cavity dimensions extracted from the referenced papers. There is exceptional agreement between the theory and the experimental results, demonstrating that regardless of cavity or sample geometry, mode splittings can be accurately predicted. This result provides experimentalists with 100\% control and predictability of cavity--magnon experiments in the planing stage, prior to cavity construction. Even more control is available if one considers combining this predictive capability with the fine tuning of $g_{cm}$ available through spintronic techniques \cite{castel19}. This is of great value if one wishes to interact with other, less manipulatable, resonant systems, or to precisely position a cavity-magnon polariton in frequency space.
Finally, this analysis reveals the limiting parameters for achieving the DSC regime in cavity--magnon experiments using YIG. Given that $g_{cm}$ will always sit along the dot--dashed line in Fig.~\ref{fig:4}(a), or the orange surface in Fig.~\ref{fig:4}(b), we can determine the parameter space in which $g_{cm}>\omega_c$. For the optimal scenario of $\eta=1$, a frequency of $\omega_c<1.72$~GHz would achieve DSC, and therefore is not prohibited. However, achieving $\eta=1$ is a difficult task. The largest value achieved in the experiments conducted here is $\sim0.04$ for the thin film sample in the CP coupled to the BM. This would require $\omega_c<2.75$~MHz, calculated from the equation for the curved intersection line in Fig.~\ref{fig:4}(b).

	
Nonetheless, $\eta$ values of 0.18 \cite{goryachev14}, 0.39 \cite{Tabuchi113} and 0.82 \cite{Flower19} have been achieved previously. With clever cavity engineering,   more achievable than ever with 3D printing technology, there is nothing preventing the DSC regime being achieved in such systems. Furthermore, by increasing the density of spins by using an alternative magnetic material, the orange surface in Fig.~\ref{fig:4}(b) will move up and the frequency restrictions of DSC become relaxed. For example, lithium ferrite has a spin density 2.13 times larger than YIG and the same magnetic moment $\mu$ \cite{Flower19,goryachev18}, therefore an experiment with $\eta=1$ would require $\omega_c<3.67$~GHz to reach DSC.

\section{\label{sec:6}Conclusion}

By utilising a novel cavity design, cavity--magnon polariton coupling strengths are analysed over a broad range of frequencies from 1--8~GHz. The experimental data enabled a proof of a universal, generalised theoretical framework fo calculating coupling rates. This theory was then also applied to a multitude of previous experimental results taken from the literature. Through an alternative definition of the form factor $\eta$, all of these experiments can be directly compared. It is shown that any cavity--magnon experiment can be correctly predicted from simulation alone, regardless of cavity or sample geometry. The theoretical equations and physical constants described in this paper should be used to define and categorise any future cavity--magnon experiment. 

The results of this research allow future experiments to be precisely engineered for exact positioning of hybridised mode frequencies for such purposes as interacting cavity--magnon polaritons with other systems. The pathway to achieving the DSC regime in these systems, something not yet achieved, is also laid out. \\


\section*{Supplementary Materials}
See supplementary material for cavity and magnon linewidth measurements, and data used to calculate $\eta$ values.

\section*{Data Availability}
The data that supports the findings of this study are available within the article (and its supplementary material).

\section*{Acknowledgements} 
\textit{This work is funded by the R\'egion Bretagne through the project OSCAR-SAD18024. This work is also part of the research program supported by the European Union through the European Regional Development Fund (ERDF) and by Ministry of Higher Education and Research, Brittany and Rennes M\'etropole, through the CPER Project SOPHIE/STIC $\&$ Ondes.}


\begin{thebibliography}{60}%
\makeatletter
\providecommand \@ifxundefined [1]{%
 \@ifx{#1\undefined}
}%
\providecommand \@ifnum [1]{%
 \ifnum #1\expandafter \@firstoftwo
 \else \expandafter \@secondoftwo
 \fi
}%
\providecommand \@ifx [1]{%
 \ifx #1\expandafter \@firstoftwo
 \else \expandafter \@secondoftwo
 \fi
}%
\providecommand \natexlab [1]{#1}%
\providecommand \enquote  [1]{``#1''}%
\providecommand \bibnamefont  [1]{#1}%
\providecommand \bibfnamefont [1]{#1}%
\providecommand \citenamefont [1]{#1}%
\providecommand \href@noop [0]{\@secondoftwo}%
\providecommand \href [0]{\begingroup \@sanitize@url \@href}%
\providecommand \@href[1]{\@@startlink{#1}\@@href}%
\providecommand \@@href[1]{\endgroup#1\@@endlink}%
\providecommand \@sanitize@url [0]{\catcode `\\12\catcode `\$12\catcode
  `\&12\catcode `\#12\catcode `\^12\catcode `\_12\catcode `\%12\relax}%
\providecommand \@@startlink[1]{}%
\providecommand \@@endlink[0]{}%
\providecommand \url  [0]{\begingroup\@sanitize@url \@url }%
\providecommand \@url [1]{\endgroup\@href {#1}{\urlprefix }}%
\providecommand \urlprefix  [0]{URL }%
\providecommand \Eprint [0]{\href }%
\providecommand \doibase [0]{http://dx.doi.org/}%
\providecommand \selectlanguage [0]{\@gobble}%
\providecommand \bibinfo  [0]{\@secondoftwo}%
\providecommand \bibfield  [0]{\@secondoftwo}%
\providecommand \translation [1]{[#1]}%
\providecommand \BibitemOpen [0]{}%
\providecommand \bibitemStop [0]{}%
\providecommand \bibitemNoStop [0]{.\EOS\space}%
\providecommand \EOS [0]{\spacefactor3000\relax}%
\providecommand \BibitemShut  [1]{\csname bibitem#1\endcsname}%
\let\auto@bib@innerbib\@empty
\bibitem [{\citenamefont {Chumak}\ \emph {et~al.}(2014)\citenamefont {Chumak},
  \citenamefont {Serga},\ and\ \citenamefont {Hillebrands}}]{Chumak14}%
  \BibitemOpen
  \bibfield  {author} {\bibinfo {author} {\bibfnamefont {A.~V.}\ \bibnamefont
  {Chumak}}, \bibinfo {author} {\bibfnamefont {A.~A.}\ \bibnamefont {Serga}}, \
  and\ \bibinfo {author} {\bibfnamefont {B.}~\bibnamefont {Hillebrands}},\
  }\href {\doibase 10.1038/ncomms5700} {\bibfield  {journal} {\bibinfo
  {journal} {Nature Communications}\ }\textbf {\bibinfo {volume} {5}},\
  \bibinfo {pages} {4700} (\bibinfo {year} {2014})}\BibitemShut {NoStop}%
\bibitem [{\citenamefont {Chumak}\ \emph {et~al.}(2015)\citenamefont {Chumak},
  \citenamefont {Vasyuchka}, \citenamefont {Serga},\ and\ \citenamefont
  {Hillebrands}}]{Chumak15}%
  \BibitemOpen
  \bibfield  {author} {\bibinfo {author} {\bibfnamefont {A.~V.}\ \bibnamefont
  {Chumak}}, \bibinfo {author} {\bibfnamefont {V.~I.}\ \bibnamefont
  {Vasyuchka}}, \bibinfo {author} {\bibfnamefont {A.~A.}\ \bibnamefont
  {Serga}}, \ and\ \bibinfo {author} {\bibfnamefont {B.}~\bibnamefont
  {Hillebrands}},\ }\href {https://doi.org/10.1038/nphys3347} {\bibfield
  {journal} {\bibinfo  {journal} {Nature Physics}\ }\textbf {\bibinfo {volume}
  {11}},\ \bibinfo {pages} {453 EP } (\bibinfo {year} {2015})}\BibitemShut
  {NoStop}%
\bibitem [{\citenamefont {Tanji}\ \emph {et~al.}(2009)\citenamefont {Tanji},
  \citenamefont {Ghosh}, \citenamefont {Simon}, \citenamefont {Bloom},\ and\
  \citenamefont {Vuleti\ifmmode~\acute{c}\else \'{c}\fi{}}}]{Tanji09}%
  \BibitemOpen
  \bibfield  {author} {\bibinfo {author} {\bibfnamefont {H.}~\bibnamefont
  {Tanji}}, \bibinfo {author} {\bibfnamefont {S.}~\bibnamefont {Ghosh}},
  \bibinfo {author} {\bibfnamefont {J.}~\bibnamefont {Simon}}, \bibinfo
  {author} {\bibfnamefont {B.}~\bibnamefont {Bloom}}, \ and\ \bibinfo {author}
  {\bibfnamefont {V.}~\bibnamefont {Vuleti\ifmmode~\acute{c}\else
  \'{c}\fi{}}},\ }\href {\doibase 10.1103/PhysRevLett.103.043601} {\bibfield
  {journal} {\bibinfo  {journal} {Phys. Rev. Lett.}\ }\textbf {\bibinfo
  {volume} {103}},\ \bibinfo {pages} {043601} (\bibinfo {year}
  {2009})}\BibitemShut {NoStop}%
\bibitem [{\citenamefont {Zhang}\ \emph {et~al.}(2015)\citenamefont {Zhang},
  \citenamefont {Zou}, \citenamefont {Zhu}, \citenamefont {Marquardt},
  \citenamefont {Jiang},\ and\ \citenamefont {Tang}}]{Zhang15}%
  \BibitemOpen
  \bibfield  {author} {\bibinfo {author} {\bibfnamefont {X.}~\bibnamefont
  {Zhang}}, \bibinfo {author} {\bibfnamefont {C.-L.}\ \bibnamefont {Zou}},
  \bibinfo {author} {\bibfnamefont {N.}~\bibnamefont {Zhu}}, \bibinfo {author}
  {\bibfnamefont {F.}~\bibnamefont {Marquardt}}, \bibinfo {author}
  {\bibfnamefont {L.}~\bibnamefont {Jiang}}, \ and\ \bibinfo {author}
  {\bibfnamefont {H.~X.}\ \bibnamefont {Tang}},\ }\href {\doibase
  10.1038/ncomms9914} {\bibfield  {journal} {\bibinfo  {journal} {Nature
  Communications}\ }\textbf {\bibinfo {volume} {6}},\ \bibinfo {pages} {8914}
  (\bibinfo {year} {2015})}\BibitemShut {NoStop}%
\bibitem [{\citenamefont {Wesenberg}\ \emph {et~al.}(2009)\citenamefont
  {Wesenberg}, \citenamefont {Ardavan}, \citenamefont {Briggs}, \citenamefont
  {Morton}, \citenamefont {Schoelkopf}, \citenamefont {Schuster},\ and\
  \citenamefont {M\o{}lmer}}]{wesenberg}%
  \BibitemOpen
  \bibfield  {author} {\bibinfo {author} {\bibfnamefont {J.~H.}\ \bibnamefont
  {Wesenberg}}, \bibinfo {author} {\bibfnamefont {A.}~\bibnamefont {Ardavan}},
  \bibinfo {author} {\bibfnamefont {G.~A.~D.}\ \bibnamefont {Briggs}}, \bibinfo
  {author} {\bibfnamefont {J.~J.~L.}\ \bibnamefont {Morton}}, \bibinfo {author}
  {\bibfnamefont {R.~J.}\ \bibnamefont {Schoelkopf}}, \bibinfo {author}
  {\bibfnamefont {D.~I.}\ \bibnamefont {Schuster}}, \ and\ \bibinfo {author}
  {\bibfnamefont {K.}~\bibnamefont {M\o{}lmer}},\ }\href {\doibase
  10.1103/PhysRevLett.103.070502} {\bibfield  {journal} {\bibinfo  {journal}
  {Phys. Rev. Lett.}\ }\textbf {\bibinfo {volume} {103}},\ \bibinfo {pages}
  {070502} (\bibinfo {year} {2009})}\BibitemShut {NoStop}%
\bibitem [{\citenamefont {Tabuchi}\ \emph {et~al.}(2015)\citenamefont
  {Tabuchi}, \citenamefont {Ishino}, \citenamefont {Noguchi}, \citenamefont
  {Ishikawa}, \citenamefont {Yamazaki}, \citenamefont {Usami},\ and\
  \citenamefont {Nakamura}}]{Tabuchi405}%
  \BibitemOpen
  \bibfield  {author} {\bibinfo {author} {\bibfnamefont {Y.}~\bibnamefont
  {Tabuchi}}, \bibinfo {author} {\bibfnamefont {S.}~\bibnamefont {Ishino}},
  \bibinfo {author} {\bibfnamefont {A.}~\bibnamefont {Noguchi}}, \bibinfo
  {author} {\bibfnamefont {T.}~\bibnamefont {Ishikawa}}, \bibinfo {author}
  {\bibfnamefont {R.}~\bibnamefont {Yamazaki}}, \bibinfo {author}
  {\bibfnamefont {K.}~\bibnamefont {Usami}}, \ and\ \bibinfo {author}
  {\bibfnamefont {Y.}~\bibnamefont {Nakamura}},\ }\href {\doibase
  10.1126/science.aaa3693} {\bibfield  {journal} {\bibinfo  {journal}
  {Science}\ }\textbf {\bibinfo {volume} {349}},\ \bibinfo {pages} {405}
  (\bibinfo {year} {2015})}\BibitemShut {NoStop}%
\bibitem [{\citenamefont {Tabuchi}\ \emph {et~al.}(2016)\citenamefont
  {Tabuchi}, \citenamefont {Ishino}, \citenamefont {Noguchi}, \citenamefont
  {Ishikawa}, \citenamefont {Yamazaki}, \citenamefont {Usami},\ and\
  \citenamefont {Nakamura}}]{Tabuchi16}%
  \BibitemOpen
  \bibfield  {author} {\bibinfo {author} {\bibfnamefont {Y.}~\bibnamefont
  {Tabuchi}}, \bibinfo {author} {\bibfnamefont {S.}~\bibnamefont {Ishino}},
  \bibinfo {author} {\bibfnamefont {A.}~\bibnamefont {Noguchi}}, \bibinfo
  {author} {\bibfnamefont {T.}~\bibnamefont {Ishikawa}}, \bibinfo {author}
  {\bibfnamefont {R.}~\bibnamefont {Yamazaki}}, \bibinfo {author}
  {\bibfnamefont {K.}~\bibnamefont {Usami}}, \ and\ \bibinfo {author}
  {\bibfnamefont {Y.}~\bibnamefont {Nakamura}},\ }\href {\doibase
  https://doi.org/10.1016/j.crhy.2016.07.009} {\bibfield  {journal} {\bibinfo
  {journal} {Comptes Rendus Physique}\ }\textbf {\bibinfo {volume} {17}},\
  \bibinfo {pages} {729 } (\bibinfo {year} {2016})}\BibitemShut {NoStop}%
\bibitem [{\citenamefont {Lachance-Quirion}\ \emph {et~al.}(2019)\citenamefont
  {Lachance-Quirion}, \citenamefont {Tabuchi}, \citenamefont {Gloppe},
  \citenamefont {Usami},\ and\ \citenamefont {Nakamura}}]{Nakamura_Review}%
  \BibitemOpen
  \bibfield  {author} {\bibinfo {author} {\bibfnamefont {D.}~\bibnamefont
  {Lachance-Quirion}}, \bibinfo {author} {\bibfnamefont {Y.}~\bibnamefont
  {Tabuchi}}, \bibinfo {author} {\bibfnamefont {A.}~\bibnamefont {Gloppe}},
  \bibinfo {author} {\bibfnamefont {K.}~\bibnamefont {Usami}}, \ and\ \bibinfo
  {author} {\bibfnamefont {Y.}~\bibnamefont {Nakamura}},\ }\href {\doibase
  10.7567/1882-0786/ab248d} {\bibfield  {journal} {\bibinfo  {journal} {Applied
  Physics Express}\ }\textbf {\bibinfo {volume} {12}},\ \bibinfo {pages}
  {070101} (\bibinfo {year} {2019})}\BibitemShut {NoStop}%
\bibitem [{\citenamefont {Zhang}\ \emph
  {et~al.}(2016{\natexlab{a}})\citenamefont {Zhang}, \citenamefont {Zhu},
  \citenamefont {Zou},\ and\ \citenamefont {Tang}}]{Zhang2}%
  \BibitemOpen
  \bibfield  {author} {\bibinfo {author} {\bibfnamefont {X.}~\bibnamefont
  {Zhang}}, \bibinfo {author} {\bibfnamefont {N.}~\bibnamefont {Zhu}}, \bibinfo
  {author} {\bibfnamefont {C.-L.}\ \bibnamefont {Zou}}, \ and\ \bibinfo
  {author} {\bibfnamefont {H.~X.}\ \bibnamefont {Tang}},\ }\href {\doibase
  10.1103/PhysRevLett.117.123605} {\bibfield  {journal} {\bibinfo  {journal}
  {Phys. Rev. Lett.}\ }\textbf {\bibinfo {volume} {117}},\ \bibinfo {pages}
  {123605} (\bibinfo {year} {2016}{\natexlab{a}})}\BibitemShut {NoStop}%
\bibitem [{\citenamefont {Osada}\ \emph {et~al.}(2016)\citenamefont {Osada},
  \citenamefont {Hisatomi}, \citenamefont {Noguchi}, \citenamefont {Tabuchi},
  \citenamefont {Yamazaki}, \citenamefont {Usami}, \citenamefont {Sadgrove},
  \citenamefont {Yalla}, \citenamefont {Nomura},\ and\ \citenamefont
  {Nakamura}}]{Osada}%
  \BibitemOpen
  \bibfield  {author} {\bibinfo {author} {\bibfnamefont {A.}~\bibnamefont
  {Osada}}, \bibinfo {author} {\bibfnamefont {R.}~\bibnamefont {Hisatomi}},
  \bibinfo {author} {\bibfnamefont {A.}~\bibnamefont {Noguchi}}, \bibinfo
  {author} {\bibfnamefont {Y.}~\bibnamefont {Tabuchi}}, \bibinfo {author}
  {\bibfnamefont {R.}~\bibnamefont {Yamazaki}}, \bibinfo {author}
  {\bibfnamefont {K.}~\bibnamefont {Usami}}, \bibinfo {author} {\bibfnamefont
  {M.}~\bibnamefont {Sadgrove}}, \bibinfo {author} {\bibfnamefont
  {R.}~\bibnamefont {Yalla}}, \bibinfo {author} {\bibfnamefont
  {M.}~\bibnamefont {Nomura}}, \ and\ \bibinfo {author} {\bibfnamefont
  {Y.}~\bibnamefont {Nakamura}},\ }\href {\doibase
  10.1103/PhysRevLett.116.223601} {\bibfield  {journal} {\bibinfo  {journal}
  {Phys. Rev. Lett.}\ }\textbf {\bibinfo {volume} {116}},\ \bibinfo {pages}
  {223601} (\bibinfo {year} {2016})}\BibitemShut {NoStop}%
\bibitem [{\citenamefont {Shen}\ and\ \citenamefont
  {Bloembergen}(1966)}]{Shen}%
  \BibitemOpen
  \bibfield  {author} {\bibinfo {author} {\bibfnamefont {Y.~R.}\ \bibnamefont
  {Shen}}\ and\ \bibinfo {author} {\bibfnamefont {N.}~\bibnamefont
  {Bloembergen}},\ }\href {\doibase 10.1103/PhysRev.143.372} {\bibfield
  {journal} {\bibinfo  {journal} {Phys. Rev.}\ }\textbf {\bibinfo {volume}
  {143}},\ \bibinfo {pages} {372} (\bibinfo {year} {1966})}\BibitemShut
  {NoStop}%
\bibitem [{\citenamefont {Demokritov}\ \emph {et~al.}(2001)\citenamefont
  {Demokritov}, \citenamefont {Hillebrands},\ and\ \citenamefont
  {Slavin}}]{Demokritov}%
  \BibitemOpen
  \bibfield  {author} {\bibinfo {author} {\bibfnamefont {S.}~\bibnamefont
  {Demokritov}}, \bibinfo {author} {\bibfnamefont {B.}~\bibnamefont
  {Hillebrands}}, \ and\ \bibinfo {author} {\bibfnamefont {A.}~\bibnamefont
  {Slavin}},\ }\href {\doibase http://dx.doi.org/10.1016/S0370-1573(00)00116-2}
  {\bibfield  {journal} {\bibinfo  {journal} {Physics Reports}\ }\textbf
  {\bibinfo {volume} {348}},\ \bibinfo {pages} {441 } (\bibinfo {year}
  {2001})}\BibitemShut {NoStop}%
\bibitem [{\citenamefont {Zhang}\ \emph
  {et~al.}(2016{\natexlab{b}})\citenamefont {Zhang}, \citenamefont {Zou},
  \citenamefont {Jiang},\ and\ \citenamefont {Tang}}]{Zhang3}%
  \BibitemOpen
  \bibfield  {author} {\bibinfo {author} {\bibfnamefont {X.}~\bibnamefont
  {Zhang}}, \bibinfo {author} {\bibfnamefont {C.-L.}\ \bibnamefont {Zou}},
  \bibinfo {author} {\bibfnamefont {L.}~\bibnamefont {Jiang}}, \ and\ \bibinfo
  {author} {\bibfnamefont {H.~X.}\ \bibnamefont {Tang}},\ }\href {\doibase
  10.1126/sciadv.1501286} {\bibfield  {journal} {\bibinfo  {journal} {Science
  Advances}\ }\textbf {\bibinfo {volume} {2}} (\bibinfo {year}
  {2016}{\natexlab{b}}),\ 10.1126/sciadv.1501286}\BibitemShut {NoStop}%
\bibitem [{\citenamefont {Frisk~Kockum}\ \emph {et~al.}(2019)\citenamefont
  {Frisk~Kockum}, \citenamefont {Miranowicz}, \citenamefont {De~Liberato},
  \citenamefont {Savasta},\ and\ \citenamefont {Nori}}]{Liberato2}%
  \BibitemOpen
  \bibfield  {author} {\bibinfo {author} {\bibfnamefont {A.}~\bibnamefont
  {Frisk~Kockum}}, \bibinfo {author} {\bibfnamefont {A.}~\bibnamefont
  {Miranowicz}}, \bibinfo {author} {\bibfnamefont {S.}~\bibnamefont
  {De~Liberato}}, \bibinfo {author} {\bibfnamefont {S.}~\bibnamefont
  {Savasta}}, \ and\ \bibinfo {author} {\bibfnamefont {F.}~\bibnamefont
  {Nori}},\ }\href {\doibase 10.1038/s42254-018-0006-2} {\bibfield  {journal}
  {\bibinfo  {journal} {Nature Reviews Physics}\ }\textbf {\bibinfo {volume}
  {1}},\ \bibinfo {pages} {19} (\bibinfo {year} {2019})}\BibitemShut {NoStop}%
\bibitem [{\citenamefont {Zare~Rameshti}\ \emph {et~al.}(2015)\citenamefont
  {Zare~Rameshti}, \citenamefont {Cao},\ and\ \citenamefont
  {Bauer}}]{Rameshti2}%
  \BibitemOpen
  \bibfield  {author} {\bibinfo {author} {\bibfnamefont {B.}~\bibnamefont
  {Zare~Rameshti}}, \bibinfo {author} {\bibfnamefont {Y.}~\bibnamefont {Cao}},
  \ and\ \bibinfo {author} {\bibfnamefont {G.~E.~W.}\ \bibnamefont {Bauer}},\
  }\href {\doibase 10.1103/PhysRevB.91.214430} {\bibfield  {journal} {\bibinfo
  {journal} {Phys. Rev. B}\ }\textbf {\bibinfo {volume} {91}},\ \bibinfo
  {pages} {214430} (\bibinfo {year} {2015})}\BibitemShut {NoStop}%
\bibitem [{\citenamefont {Flower}\ \emph {et~al.}(2019)\citenamefont {Flower},
  \citenamefont {Goryachev}, \citenamefont {Bourhill},\ and\ \citenamefont
  {Tobar}}]{Flower19}%
  \BibitemOpen
  \bibfield  {author} {\bibinfo {author} {\bibfnamefont {G.}~\bibnamefont
  {Flower}}, \bibinfo {author} {\bibfnamefont {M.}~\bibnamefont {Goryachev}},
  \bibinfo {author} {\bibfnamefont {J.}~\bibnamefont {Bourhill}}, \ and\
  \bibinfo {author} {\bibfnamefont {M.~E.}\ \bibnamefont {Tobar}},\ }\href
  {\doibase 10.1088/1367-2630/ab3e1c} {\bibfield  {journal} {\bibinfo
  {journal} {New Journal of Physics}\ }\textbf {\bibinfo {volume} {21}},\
  \bibinfo {pages} {095004} (\bibinfo {year} {2019})}\BibitemShut {NoStop}%
\bibitem [{\citenamefont {Bourhill}\ \emph {et~al.}(2016)\citenamefont
  {Bourhill}, \citenamefont {Kostylev}, \citenamefont {Goryachev},
  \citenamefont {Creedon},\ and\ \citenamefont {Tobar}}]{Bourhill16}%
  \BibitemOpen
  \bibfield  {author} {\bibinfo {author} {\bibfnamefont {J.}~\bibnamefont
  {Bourhill}}, \bibinfo {author} {\bibfnamefont {N.}~\bibnamefont {Kostylev}},
  \bibinfo {author} {\bibfnamefont {M.}~\bibnamefont {Goryachev}}, \bibinfo
  {author} {\bibfnamefont {D.~L.}\ \bibnamefont {Creedon}}, \ and\ \bibinfo
  {author} {\bibfnamefont {M.~E.}\ \bibnamefont {Tobar}},\ }\href {\doibase
  10.1103/PhysRevB.93.144420} {\bibfield  {journal} {\bibinfo  {journal} {Phys.
  Rev. B}\ }\textbf {\bibinfo {volume} {93}},\ \bibinfo {pages} {144420}
  (\bibinfo {year} {2016})}\BibitemShut {NoStop}%
\bibitem [{\citenamefont {Goryachev}\ \emph {et~al.}(2014)\citenamefont
  {Goryachev}, \citenamefont {Farr}, \citenamefont {Creedon}, \citenamefont
  {Fan}, \citenamefont {Kostylev},\ and\ \citenamefont {Tobar}}]{goryachev14}%
  \BibitemOpen
  \bibfield  {author} {\bibinfo {author} {\bibfnamefont {M.}~\bibnamefont
  {Goryachev}}, \bibinfo {author} {\bibfnamefont {W.~G.}\ \bibnamefont {Farr}},
  \bibinfo {author} {\bibfnamefont {D.~L.}\ \bibnamefont {Creedon}}, \bibinfo
  {author} {\bibfnamefont {Y.}~\bibnamefont {Fan}}, \bibinfo {author}
  {\bibfnamefont {M.}~\bibnamefont {Kostylev}}, \ and\ \bibinfo {author}
  {\bibfnamefont {M.~E.}\ \bibnamefont {Tobar}},\ }\href {\doibase
  10.1103/PhysRevApplied.2.054002} {\bibfield  {journal} {\bibinfo  {journal}
  {Phys. Rev. Applied}\ }\textbf {\bibinfo {volume} {2}},\ \bibinfo {pages}
  {054002} (\bibinfo {year} {2014})}\BibitemShut {NoStop}%
\bibitem [{\citenamefont {Zhang}\ \emph {et~al.}(2014)\citenamefont {Zhang},
  \citenamefont {Zou}, \citenamefont {Jiang},\ and\ \citenamefont
  {Tang}}]{Zhang1}%
  \BibitemOpen
  \bibfield  {author} {\bibinfo {author} {\bibfnamefont {X.}~\bibnamefont
  {Zhang}}, \bibinfo {author} {\bibfnamefont {C.-L.}\ \bibnamefont {Zou}},
  \bibinfo {author} {\bibfnamefont {L.}~\bibnamefont {Jiang}}, \ and\ \bibinfo
  {author} {\bibfnamefont {H.~X.}\ \bibnamefont {Tang}},\ }\href {\doibase
  10.1103/PhysRevLett.113.156401} {\bibfield  {journal} {\bibinfo  {journal}
  {Phys. Rev. Lett.}\ }\textbf {\bibinfo {volume} {113}},\ \bibinfo {pages}
  {156401} (\bibinfo {year} {2014})}\BibitemShut {NoStop}%
\bibitem [{\citenamefont {Auer}\ and\ \citenamefont {Burkard}(2012)}]{Auer}%
  \BibitemOpen
  \bibfield  {author} {\bibinfo {author} {\bibfnamefont {A.}~\bibnamefont
  {Auer}}\ and\ \bibinfo {author} {\bibfnamefont {G.}~\bibnamefont {Burkard}},\
  }\href {\doibase 10.1103/PhysRevB.85.235140} {\bibfield  {journal} {\bibinfo
  {journal} {Phys. Rev. B}\ }\textbf {\bibinfo {volume} {85}},\ \bibinfo
  {pages} {235140} (\bibinfo {year} {2012})}\BibitemShut {NoStop}%
\bibitem [{\citenamefont {De~Liberato}\ and\ \citenamefont
  {Ciuti}(2008)}]{Liberato3}%
  \BibitemOpen
  \bibfield  {author} {\bibinfo {author} {\bibfnamefont {S.}~\bibnamefont
  {De~Liberato}}\ and\ \bibinfo {author} {\bibfnamefont {C.}~\bibnamefont
  {Ciuti}},\ }\href {\doibase 10.1103/PhysRevB.77.155321} {\bibfield  {journal}
  {\bibinfo  {journal} {Phys. Rev. B}\ }\textbf {\bibinfo {volume} {77}},\
  \bibinfo {pages} {155321} (\bibinfo {year} {2008})}\BibitemShut {NoStop}%
\bibitem [{\citenamefont {De~Liberato}(2014)}]{Liberato5}%
  \BibitemOpen
  \bibfield  {author} {\bibinfo {author} {\bibfnamefont {S.}~\bibnamefont
  {De~Liberato}},\ }\href {\doibase 10.1103/PhysRevLett.112.016401} {\bibfield
  {journal} {\bibinfo  {journal} {Phys. Rev. Lett.}\ }\textbf {\bibinfo
  {volume} {112}},\ \bibinfo {pages} {016401} (\bibinfo {year}
  {2014})}\BibitemShut {NoStop}%
\bibitem [{\citenamefont {Lambert}\ \emph {et~al.}(2004)\citenamefont
  {Lambert}, \citenamefont {Emary},\ and\ \citenamefont {Brandes}}]{Lambert1}%
  \BibitemOpen
  \bibfield  {author} {\bibinfo {author} {\bibfnamefont {N.}~\bibnamefont
  {Lambert}}, \bibinfo {author} {\bibfnamefont {C.}~\bibnamefont {Emary}}, \
  and\ \bibinfo {author} {\bibfnamefont {T.}~\bibnamefont {Brandes}},\ }\href
  {\doibase 10.1103/PhysRevLett.92.073602} {\bibfield  {journal} {\bibinfo
  {journal} {Phys. Rev. Lett.}\ }\textbf {\bibinfo {volume} {92}},\ \bibinfo
  {pages} {073602} (\bibinfo {year} {2004})}\BibitemShut {NoStop}%
\bibitem [{\citenamefont {Nataf}\ and\ \citenamefont {Ciuti}(2010)}]{Nataf1}%
  \BibitemOpen
  \bibfield  {author} {\bibinfo {author} {\bibfnamefont {P.}~\bibnamefont
  {Nataf}}\ and\ \bibinfo {author} {\bibfnamefont {C.}~\bibnamefont {Ciuti}},\
  }\href {https://doi.org/10.1038/ncomms1069} {\bibfield  {journal} {\bibinfo
  {journal} {Nature Communications}\ }\textbf {\bibinfo {volume} {1}},\
  \bibinfo {pages} {72 EP } (\bibinfo {year} {2010})},\ \bibinfo {note}
  {article}\BibitemShut {NoStop}%
\bibitem [{\citenamefont {Ciuti}\ and\ \citenamefont
  {Carusotto}(2006)}]{ciuti1}%
  \BibitemOpen
  \bibfield  {author} {\bibinfo {author} {\bibfnamefont {C.}~\bibnamefont
  {Ciuti}}\ and\ \bibinfo {author} {\bibfnamefont {I.}~\bibnamefont
  {Carusotto}},\ }\href {\doibase 10.1103/PhysRevA.74.033811} {\bibfield
  {journal} {\bibinfo  {journal} {Phys. Rev. A}\ }\textbf {\bibinfo {volume}
  {74}},\ \bibinfo {pages} {033811} (\bibinfo {year} {2006})}\BibitemShut
  {NoStop}%
\bibitem [{\citenamefont {Casanova}\ \emph {et~al.}(2010)\citenamefont
  {Casanova}, \citenamefont {Romero}, \citenamefont {Lizuain}, \citenamefont
  {Garc\'{\i}a-Ripoll},\ and\ \citenamefont {Solano}}]{DSC}%
  \BibitemOpen
  \bibfield  {author} {\bibinfo {author} {\bibfnamefont {J.}~\bibnamefont
  {Casanova}}, \bibinfo {author} {\bibfnamefont {G.}~\bibnamefont {Romero}},
  \bibinfo {author} {\bibfnamefont {I.}~\bibnamefont {Lizuain}}, \bibinfo
  {author} {\bibfnamefont {J.~J.}\ \bibnamefont {Garc\'{\i}a-Ripoll}}, \ and\
  \bibinfo {author} {\bibfnamefont {E.}~\bibnamefont {Solano}},\ }\href
  {\doibase 10.1103/PhysRevLett.105.263603} {\bibfield  {journal} {\bibinfo
  {journal} {Phys. Rev. Lett.}\ }\textbf {\bibinfo {volume} {105}},\ \bibinfo
  {pages} {263603} (\bibinfo {year} {2010})}\BibitemShut {NoStop}%
\bibitem [{\citenamefont {Barberena}\ \emph {et~al.}(2017)\citenamefont
  {Barberena}, \citenamefont {Lamata},\ and\ \citenamefont
  {Solano}}]{Barberena:2017aa}%
  \BibitemOpen
  \bibfield  {author} {\bibinfo {author} {\bibfnamefont {D.}~\bibnamefont
  {Barberena}}, \bibinfo {author} {\bibfnamefont {L.}~\bibnamefont {Lamata}}, \
  and\ \bibinfo {author} {\bibfnamefont {E.}~\bibnamefont {Solano}},\ }\href
  {\doibase 10.1038/s41598-017-09110-7} {\bibfield  {journal} {\bibinfo
  {journal} {Scientific Reports}\ }\textbf {\bibinfo {volume} {7}},\ \bibinfo
  {pages} {8774} (\bibinfo {year} {2017})}\BibitemShut {NoStop}%
\bibitem [{\citenamefont {Kajiwara}\ \emph {et~al.}(2010)\citenamefont
  {Kajiwara}, \citenamefont {Harii}, \citenamefont {Takahashi}, \citenamefont
  {Ohe}, \citenamefont {Uchida}, \citenamefont {Mizuguchi}, \citenamefont
  {Umezawa}, \citenamefont {Kawai}, \citenamefont {Ando}, \citenamefont
  {Takanashi}, \citenamefont {Maekawa},\ and\ \citenamefont
  {Saitoh}}]{Kajiwara2010}%
  \BibitemOpen
  \bibfield  {author} {\bibinfo {author} {\bibfnamefont {Y.}~\bibnamefont
  {Kajiwara}}, \bibinfo {author} {\bibfnamefont {K.}~\bibnamefont {Harii}},
  \bibinfo {author} {\bibfnamefont {S.}~\bibnamefont {Takahashi}}, \bibinfo
  {author} {\bibfnamefont {J.}~\bibnamefont {Ohe}}, \bibinfo {author}
  {\bibfnamefont {K.}~\bibnamefont {Uchida}}, \bibinfo {author} {\bibfnamefont
  {M.}~\bibnamefont {Mizuguchi}}, \bibinfo {author} {\bibfnamefont
  {H.}~\bibnamefont {Umezawa}}, \bibinfo {author} {\bibfnamefont
  {H.}~\bibnamefont {Kawai}}, \bibinfo {author} {\bibfnamefont
  {K.}~\bibnamefont {Ando}}, \bibinfo {author} {\bibfnamefont {K.}~\bibnamefont
  {Takanashi}}, \bibinfo {author} {\bibfnamefont {S.}~\bibnamefont {Maekawa}},
  \ and\ \bibinfo {author} {\bibfnamefont {E.}~\bibnamefont {Saitoh}},\ }\href
  {\doibase 10.1038/nature08876} {\bibfield  {journal} {\bibinfo  {journal}
  {Nature}\ }\textbf {\bibinfo {volume} {464}},\ \bibinfo {pages} {262}
  (\bibinfo {year} {2010})}\BibitemShut {NoStop}%
\bibitem [{\citenamefont {Wang}\ \emph {et~al.}(2014)\citenamefont {Wang},
  \citenamefont {Du}, \citenamefont {Pu}, \citenamefont {Adur}, \citenamefont
  {Hammel},\ and\ \citenamefont {Yang}}]{Wang1}%
  \BibitemOpen
  \bibfield  {author} {\bibinfo {author} {\bibfnamefont {H.~L.}\ \bibnamefont
  {Wang}}, \bibinfo {author} {\bibfnamefont {C.~H.}\ \bibnamefont {Du}},
  \bibinfo {author} {\bibfnamefont {Y.}~\bibnamefont {Pu}}, \bibinfo {author}
  {\bibfnamefont {R.}~\bibnamefont {Adur}}, \bibinfo {author} {\bibfnamefont
  {P.~C.}\ \bibnamefont {Hammel}}, \ and\ \bibinfo {author} {\bibfnamefont
  {F.~Y.}\ \bibnamefont {Yang}},\ }\href {\doibase
  10.1103/PhysRevLett.112.197201} {\bibfield  {journal} {\bibinfo  {journal}
  {Phys. Rev. Lett.}\ }\textbf {\bibinfo {volume} {112}},\ \bibinfo {pages}
  {197201} (\bibinfo {year} {2014})}\BibitemShut {NoStop}%
\bibitem [{\citenamefont {Czeschka}\ \emph {et~al.}(2011)\citenamefont
  {Czeschka}, \citenamefont {Dreher}, \citenamefont {Brandt}, \citenamefont
  {Weiler}, \citenamefont {Althammer}, \citenamefont {Imort}, \citenamefont
  {Reiss}, \citenamefont {Thomas}, \citenamefont {Schoch}, \citenamefont
  {Limmer}, \citenamefont {Huebl}, \citenamefont {Gross},\ and\ \citenamefont
  {Goennenwein}}]{czeschka}%
  \BibitemOpen
  \bibfield  {author} {\bibinfo {author} {\bibfnamefont {F.~D.}\ \bibnamefont
  {Czeschka}}, \bibinfo {author} {\bibfnamefont {L.}~\bibnamefont {Dreher}},
  \bibinfo {author} {\bibfnamefont {M.~S.}\ \bibnamefont {Brandt}}, \bibinfo
  {author} {\bibfnamefont {M.}~\bibnamefont {Weiler}}, \bibinfo {author}
  {\bibfnamefont {M.}~\bibnamefont {Althammer}}, \bibinfo {author}
  {\bibfnamefont {I.-M.}\ \bibnamefont {Imort}}, \bibinfo {author}
  {\bibfnamefont {G.}~\bibnamefont {Reiss}}, \bibinfo {author} {\bibfnamefont
  {A.}~\bibnamefont {Thomas}}, \bibinfo {author} {\bibfnamefont
  {W.}~\bibnamefont {Schoch}}, \bibinfo {author} {\bibfnamefont
  {W.}~\bibnamefont {Limmer}}, \bibinfo {author} {\bibfnamefont
  {H.}~\bibnamefont {Huebl}}, \bibinfo {author} {\bibfnamefont
  {R.}~\bibnamefont {Gross}}, \ and\ \bibinfo {author} {\bibfnamefont
  {S.~T.~B.}\ \bibnamefont {Goennenwein}},\ }\href {\doibase
  10.1103/PhysRevLett.107.046601} {\bibfield  {journal} {\bibinfo  {journal}
  {Phys. Rev. Lett.}\ }\textbf {\bibinfo {volume} {107}},\ \bibinfo {pages}
  {046601} (\bibinfo {year} {2011})}\BibitemShut {NoStop}%
\bibitem [{\citenamefont {Castel}\ \emph {et~al.}(2012)\citenamefont {Castel},
  \citenamefont {Vlietstra}, \citenamefont {Ben~Youssef},\ and\ \citenamefont
  {van Wees}}]{castel2}%
  \BibitemOpen
  \bibfield  {author} {\bibinfo {author} {\bibfnamefont {V.}~\bibnamefont
  {Castel}}, \bibinfo {author} {\bibfnamefont {N.}~\bibnamefont {Vlietstra}},
  \bibinfo {author} {\bibfnamefont {J.}~\bibnamefont {Ben~Youssef}}, \ and\
  \bibinfo {author} {\bibfnamefont {B.~J.}\ \bibnamefont {van Wees}},\ }\href
  {\doibase 10.1063/1.4754837} {\bibfield  {journal} {\bibinfo  {journal}
  {Applied Physics Letters}\ }\textbf {\bibinfo {volume} {101}},\ \bibinfo
  {pages} {132414} (\bibinfo {year} {2012})}\BibitemShut {NoStop}%
\bibitem [{\citenamefont {Maier-Flaig}\ \emph {et~al.}(2016)\citenamefont
  {Maier-Flaig}, \citenamefont {Harder}, \citenamefont {Gross}, \citenamefont
  {Huebl},\ and\ \citenamefont {Goennenwein}}]{Maier-Flaig1}%
  \BibitemOpen
  \bibfield  {author} {\bibinfo {author} {\bibfnamefont {H.}~\bibnamefont
  {Maier-Flaig}}, \bibinfo {author} {\bibfnamefont {M.}~\bibnamefont {Harder}},
  \bibinfo {author} {\bibfnamefont {R.}~\bibnamefont {Gross}}, \bibinfo
  {author} {\bibfnamefont {H.}~\bibnamefont {Huebl}}, \ and\ \bibinfo {author}
  {\bibfnamefont {S.~T.~B.}\ \bibnamefont {Goennenwein}},\ }\href {\doibase
  10.1103/PhysRevB.94.054433} {\bibfield  {journal} {\bibinfo  {journal} {Phys.
  Rev. B}\ }\textbf {\bibinfo {volume} {94}},\ \bibinfo {pages} {054433}
  (\bibinfo {year} {2016})}\BibitemShut {NoStop}%
\bibitem [{\citenamefont {Bai}\ \emph {et~al.}(2015)\citenamefont {Bai},
  \citenamefont {Harder}, \citenamefont {Chen}, \citenamefont {Fan},
  \citenamefont {Xiao},\ and\ \citenamefont {Hu}}]{Bai1}%
  \BibitemOpen
  \bibfield  {author} {\bibinfo {author} {\bibfnamefont {L.}~\bibnamefont
  {Bai}}, \bibinfo {author} {\bibfnamefont {M.}~\bibnamefont {Harder}},
  \bibinfo {author} {\bibfnamefont {Y.~P.}\ \bibnamefont {Chen}}, \bibinfo
  {author} {\bibfnamefont {X.}~\bibnamefont {Fan}}, \bibinfo {author}
  {\bibfnamefont {J.~Q.}\ \bibnamefont {Xiao}}, \ and\ \bibinfo {author}
  {\bibfnamefont {C.-M.}\ \bibnamefont {Hu}},\ }\href {\doibase
  10.1103/PhysRevLett.114.227201} {\bibfield  {journal} {\bibinfo  {journal}
  {Phys. Rev. Lett.}\ }\textbf {\bibinfo {volume} {114}},\ \bibinfo {pages}
  {227201} (\bibinfo {year} {2015})}\BibitemShut {NoStop}%
\bibitem [{\citenamefont {Cao}\ \emph {et~al.}(2015)\citenamefont {Cao},
  \citenamefont {Yan}, \citenamefont {Huebl}, \citenamefont {Goennenwein},\
  and\ \citenamefont {Bauer}}]{Cao1}%
  \BibitemOpen
  \bibfield  {author} {\bibinfo {author} {\bibfnamefont {Y.}~\bibnamefont
  {Cao}}, \bibinfo {author} {\bibfnamefont {P.}~\bibnamefont {Yan}}, \bibinfo
  {author} {\bibfnamefont {H.}~\bibnamefont {Huebl}}, \bibinfo {author}
  {\bibfnamefont {S.~T.~B.}\ \bibnamefont {Goennenwein}}, \ and\ \bibinfo
  {author} {\bibfnamefont {G.~E.~W.}\ \bibnamefont {Bauer}},\ }\href {\doibase
  10.1103/PhysRevB.91.094423} {\bibfield  {journal} {\bibinfo  {journal} {Phys.
  Rev. B}\ }\textbf {\bibinfo {volume} {91}},\ \bibinfo {pages} {094423}
  (\bibinfo {year} {2015})}\BibitemShut {NoStop}%
\bibitem [{\citenamefont {Zhang}\ \emph
  {et~al.}(2016{\natexlab{c}})\citenamefont {Zhang}, \citenamefont {Zou},
  \citenamefont {Jiang},\ and\ \citenamefont {Tang}}]{Zhang4}%
  \BibitemOpen
  \bibfield  {author} {\bibinfo {author} {\bibfnamefont {X.}~\bibnamefont
  {Zhang}}, \bibinfo {author} {\bibfnamefont {C.}~\bibnamefont {Zou}}, \bibinfo
  {author} {\bibfnamefont {L.}~\bibnamefont {Jiang}}, \ and\ \bibinfo {author}
  {\bibfnamefont {H.~X.}\ \bibnamefont {Tang}},\ }\href {\doibase
  10.1063/1.4939134} {\bibfield  {journal} {\bibinfo  {journal} {Journal of
  Applied Physics}\ }\textbf {\bibinfo {volume} {119}},\ \bibinfo {pages}
  {023905} (\bibinfo {year} {2016}{\natexlab{c}})},\ \Eprint
  {http://arxiv.org/abs/https://doi.org/10.1063/1.4939134}
  {https://doi.org/10.1063/1.4939134} \BibitemShut {NoStop}%
\bibitem [{\citenamefont {Tabuchi}\ \emph {et~al.}(2014)\citenamefont
  {Tabuchi}, \citenamefont {Ishino}, \citenamefont {Ishikawa}, \citenamefont
  {Yamazaki}, \citenamefont {Usami},\ and\ \citenamefont
  {Nakamura}}]{Tabuchi113}%
  \BibitemOpen
  \bibfield  {author} {\bibinfo {author} {\bibfnamefont {Y.}~\bibnamefont
  {Tabuchi}}, \bibinfo {author} {\bibfnamefont {S.}~\bibnamefont {Ishino}},
  \bibinfo {author} {\bibfnamefont {T.}~\bibnamefont {Ishikawa}}, \bibinfo
  {author} {\bibfnamefont {R.}~\bibnamefont {Yamazaki}}, \bibinfo {author}
  {\bibfnamefont {K.}~\bibnamefont {Usami}}, \ and\ \bibinfo {author}
  {\bibfnamefont {Y.}~\bibnamefont {Nakamura}},\ }\href {\doibase
  10.1103/PhysRevLett.113.083603} {\bibfield  {journal} {\bibinfo  {journal}
  {Phys. Rev. Lett.}\ }\textbf {\bibinfo {volume} {113}},\ \bibinfo {pages}
  {083603} (\bibinfo {year} {2014})}\BibitemShut {NoStop}%
\bibitem [{\citenamefont {Zare~Rameshti}\ and\ \citenamefont
  {Bauer}(2018)}]{Rameshti}%
  \BibitemOpen
  \bibfield  {author} {\bibinfo {author} {\bibfnamefont {B.}~\bibnamefont
  {Zare~Rameshti}}\ and\ \bibinfo {author} {\bibfnamefont {G.~E.~W.}\
  \bibnamefont {Bauer}},\ }\href {\doibase 10.1103/PhysRevB.97.014419}
  {\bibfield  {journal} {\bibinfo  {journal} {Phys. Rev. B}\ }\textbf {\bibinfo
  {volume} {97}},\ \bibinfo {pages} {014419} (\bibinfo {year}
  {2018})}\BibitemShut {NoStop}%
\bibitem [{\citenamefont {Bai}\ \emph {et~al.}(2017)\citenamefont {Bai},
  \citenamefont {Harder}, \citenamefont {Hyde}, \citenamefont {Zhang},
  \citenamefont {Hu}, \citenamefont {Chen},\ and\ \citenamefont
  {Xiao}}]{PhysRevLett.118.217201}%
  \BibitemOpen
  \bibfield  {author} {\bibinfo {author} {\bibfnamefont {L.}~\bibnamefont
  {Bai}}, \bibinfo {author} {\bibfnamefont {M.}~\bibnamefont {Harder}},
  \bibinfo {author} {\bibfnamefont {P.}~\bibnamefont {Hyde}}, \bibinfo {author}
  {\bibfnamefont {Z.}~\bibnamefont {Zhang}}, \bibinfo {author} {\bibfnamefont
  {C.-M.}\ \bibnamefont {Hu}}, \bibinfo {author} {\bibfnamefont {Y.~P.}\
  \bibnamefont {Chen}}, \ and\ \bibinfo {author} {\bibfnamefont {J.~Q.}\
  \bibnamefont {Xiao}},\ }\href {\doibase 10.1103/PhysRevLett.118.217201}
  {\bibfield  {journal} {\bibinfo  {journal} {Phys. Rev. Lett.}\ }\textbf
  {\bibinfo {volume} {118}},\ \bibinfo {pages} {217201} (\bibinfo {year}
  {2017})}\BibitemShut {NoStop}%
\bibitem [{\citenamefont {Xu}\ \emph {et~al.}(2019)\citenamefont {Xu},
  \citenamefont {Rao}, \citenamefont {Gui}, \citenamefont {Jin},\ and\
  \citenamefont {Hu}}]{PhysRevB.100.094415}%
  \BibitemOpen
  \bibfield  {author} {\bibinfo {author} {\bibfnamefont {P.-C.}\ \bibnamefont
  {Xu}}, \bibinfo {author} {\bibfnamefont {J.~W.}\ \bibnamefont {Rao}},
  \bibinfo {author} {\bibfnamefont {Y.~S.}\ \bibnamefont {Gui}}, \bibinfo
  {author} {\bibfnamefont {X.}~\bibnamefont {Jin}}, \ and\ \bibinfo {author}
  {\bibfnamefont {C.-M.}\ \bibnamefont {Hu}},\ }\href {\doibase
  10.1103/PhysRevB.100.094415} {\bibfield  {journal} {\bibinfo  {journal}
  {Phys. Rev. B}\ }\textbf {\bibinfo {volume} {100}},\ \bibinfo {pages}
  {094415} (\bibinfo {year} {2019})}\BibitemShut {NoStop}%
\bibitem [{\citenamefont {Lambert}\ \emph {et~al.}(2016)\citenamefont
  {Lambert}, \citenamefont {Haigh}, \citenamefont {Langenfeld}, \citenamefont
  {Doherty},\ and\ \citenamefont {Ferguson}}]{PhysRevA.93.021803}%
  \BibitemOpen
  \bibfield  {author} {\bibinfo {author} {\bibfnamefont {N.~J.}\ \bibnamefont
  {Lambert}}, \bibinfo {author} {\bibfnamefont {J.~A.}\ \bibnamefont {Haigh}},
  \bibinfo {author} {\bibfnamefont {S.}~\bibnamefont {Langenfeld}}, \bibinfo
  {author} {\bibfnamefont {A.~C.}\ \bibnamefont {Doherty}}, \ and\ \bibinfo
  {author} {\bibfnamefont {A.~J.}\ \bibnamefont {Ferguson}},\ }\href {\doibase
  10.1103/PhysRevA.93.021803} {\bibfield  {journal} {\bibinfo  {journal} {Phys.
  Rev. A}\ }\textbf {\bibinfo {volume} {93}},\ \bibinfo {pages} {021803}
  (\bibinfo {year} {2016})}\BibitemShut {NoStop}%
\bibitem [{\citenamefont {Weichselbaumer}\ \emph {et~al.}(2019)\citenamefont
  {Weichselbaumer}, \citenamefont {Natzkin}, \citenamefont {Zollitsch},
  \citenamefont {Weiler}, \citenamefont {Gross},\ and\ \citenamefont
  {Huebl}}]{weichselbaumer}%
  \BibitemOpen
  \bibfield  {author} {\bibinfo {author} {\bibfnamefont {S.}~\bibnamefont
  {Weichselbaumer}}, \bibinfo {author} {\bibfnamefont {P.}~\bibnamefont
  {Natzkin}}, \bibinfo {author} {\bibfnamefont {C.~W.}\ \bibnamefont
  {Zollitsch}}, \bibinfo {author} {\bibfnamefont {M.}~\bibnamefont {Weiler}},
  \bibinfo {author} {\bibfnamefont {R.}~\bibnamefont {Gross}}, \ and\ \bibinfo
  {author} {\bibfnamefont {H.}~\bibnamefont {Huebl}},\ }\href {\doibase
  10.1103/PhysRevApplied.12.024021} {\bibfield  {journal} {\bibinfo  {journal}
  {Phys. Rev. Applied}\ }\textbf {\bibinfo {volume} {12}},\ \bibinfo {pages}
  {024021} (\bibinfo {year} {2019})}\BibitemShut {NoStop}%
\bibitem [{\citenamefont {{Liu}}\ \emph {et~al.}(2019)\citenamefont {{Liu}},
  \citenamefont {{Xiong}},\ and\ \citenamefont {{Wu}}}]{Liu}%
  \BibitemOpen
  \bibfield  {author} {\bibinfo {author} {\bibfnamefont {Z.}~\bibnamefont
  {{Liu}}}, \bibinfo {author} {\bibfnamefont {H.}~\bibnamefont {{Xiong}}}, \
  and\ \bibinfo {author} {\bibfnamefont {Y.}~\bibnamefont {{Wu}}},\ }\href
  {\doibase 10.1109/ACCESS.2019.2913788} {\bibfield  {journal} {\bibinfo
  {journal} {IEEE Access}\ }\textbf {\bibinfo {volume} {7}},\ \bibinfo {pages}
  {57047} (\bibinfo {year} {2019})}\BibitemShut {NoStop}%
\bibitem [{\citenamefont {Boventer}\ \emph
  {et~al.}(2018{\natexlab{a}})\citenamefont {Boventer}, \citenamefont
  {Pfirrmann}, \citenamefont {Krause}, \citenamefont {Sch\"on}, \citenamefont
  {Kl\"aui},\ and\ \citenamefont {Weides}}]{Boventer}%
  \BibitemOpen
  \bibfield  {author} {\bibinfo {author} {\bibfnamefont {I.}~\bibnamefont
  {Boventer}}, \bibinfo {author} {\bibfnamefont {M.}~\bibnamefont {Pfirrmann}},
  \bibinfo {author} {\bibfnamefont {J.}~\bibnamefont {Krause}}, \bibinfo
  {author} {\bibfnamefont {Y.}~\bibnamefont {Sch\"on}}, \bibinfo {author}
  {\bibfnamefont {M.}~\bibnamefont {Kl\"aui}}, \ and\ \bibinfo {author}
  {\bibfnamefont {M.}~\bibnamefont {Weides}},\ }\href {\doibase
  10.1103/PhysRevB.97.184420} {\bibfield  {journal} {\bibinfo  {journal} {Phys.
  Rev. B}\ }\textbf {\bibinfo {volume} {97}},\ \bibinfo {pages} {184420}
  (\bibinfo {year} {2018}{\natexlab{a}})}\BibitemShut {NoStop}%
\bibitem [{\citenamefont {Tavis}\ and\ \citenamefont {Cummings}(1968)}]{TC}%
  \BibitemOpen
  \bibfield  {author} {\bibinfo {author} {\bibfnamefont {M.}~\bibnamefont
  {Tavis}}\ and\ \bibinfo {author} {\bibfnamefont {F.~W.}\ \bibnamefont
  {Cummings}},\ }\href {\doibase 10.1103/PhysRev.170.379} {\bibfield  {journal}
  {\bibinfo  {journal} {Phys. Rev.}\ }\textbf {\bibinfo {volume} {170}},\
  \bibinfo {pages} {379} (\bibinfo {year} {1968})}\BibitemShut {NoStop}%
\bibitem [{\citenamefont {Fink}\ \emph {et~al.}(2009)\citenamefont {Fink},
  \citenamefont {Bianchetti}, \citenamefont {Baur}, \citenamefont {G\"oppl},
  \citenamefont {Steffen}, \citenamefont {Filipp}, \citenamefont {Leek},
  \citenamefont {Blais},\ and\ \citenamefont {Wallraff}}]{Fink}%
  \BibitemOpen
  \bibfield  {author} {\bibinfo {author} {\bibfnamefont {J.~M.}\ \bibnamefont
  {Fink}}, \bibinfo {author} {\bibfnamefont {R.}~\bibnamefont {Bianchetti}},
  \bibinfo {author} {\bibfnamefont {M.}~\bibnamefont {Baur}}, \bibinfo {author}
  {\bibfnamefont {M.}~\bibnamefont {G\"oppl}}, \bibinfo {author} {\bibfnamefont
  {L.}~\bibnamefont {Steffen}}, \bibinfo {author} {\bibfnamefont
  {S.}~\bibnamefont {Filipp}}, \bibinfo {author} {\bibfnamefont {P.~J.}\
  \bibnamefont {Leek}}, \bibinfo {author} {\bibfnamefont {A.}~\bibnamefont
  {Blais}}, \ and\ \bibinfo {author} {\bibfnamefont {A.}~\bibnamefont
  {Wallraff}},\ }\href {\doibase 10.1103/PhysRevLett.103.083601} {\bibfield
  {journal} {\bibinfo  {journal} {Phys. Rev. Lett.}\ }\textbf {\bibinfo
  {volume} {103}},\ \bibinfo {pages} {083601} (\bibinfo {year}
  {2009})}\BibitemShut {NoStop}%
\bibitem [{\citenamefont {Huebl}\ \emph {et~al.}(2013)\citenamefont {Huebl},
  \citenamefont {Zollitsch}, \citenamefont {Lotze}, \citenamefont {Hocke},
  \citenamefont {Greifenstein}, \citenamefont {Marx}, \citenamefont {Gross},\
  and\ \citenamefont {Goennenwein}}]{Huebl}%
  \BibitemOpen
  \bibfield  {author} {\bibinfo {author} {\bibfnamefont {H.}~\bibnamefont
  {Huebl}}, \bibinfo {author} {\bibfnamefont {C.~W.}\ \bibnamefont
  {Zollitsch}}, \bibinfo {author} {\bibfnamefont {J.}~\bibnamefont {Lotze}},
  \bibinfo {author} {\bibfnamefont {F.}~\bibnamefont {Hocke}}, \bibinfo
  {author} {\bibfnamefont {M.}~\bibnamefont {Greifenstein}}, \bibinfo {author}
  {\bibfnamefont {A.}~\bibnamefont {Marx}}, \bibinfo {author} {\bibfnamefont
  {R.}~\bibnamefont {Gross}}, \ and\ \bibinfo {author} {\bibfnamefont
  {S.~T.~B.}\ \bibnamefont {Goennenwein}},\ }\href {\doibase
  10.1103/PhysRevLett.111.127003} {\bibfield  {journal} {\bibinfo  {journal}
  {Phys. Rev. Lett.}\ }\textbf {\bibinfo {volume} {111}},\ \bibinfo {pages}
  {127003} (\bibinfo {year} {2013})}\BibitemShut {NoStop}%
\bibitem [{\citenamefont {Soykal}\ and\ \citenamefont
  {Flatt\'e}(2010)}]{Soykal}%
  \BibitemOpen
  \bibfield  {author} {\bibinfo {author} {\bibfnamefont {O.~O.}\ \bibnamefont
  {Soykal}}\ and\ \bibinfo {author} {\bibfnamefont {M.~E.}\ \bibnamefont
  {Flatt\'e}},\ }\href {\doibase 10.1103/PhysRevLett.104.077202} {\bibfield
  {journal} {\bibinfo  {journal} {Phys. Rev. Lett.}\ }\textbf {\bibinfo
  {volume} {104}},\ \bibinfo {pages} {077202} (\bibinfo {year}
  {2010})}\BibitemShut {NoStop}%
\bibitem [{\citenamefont {Gilleo}\ and\ \citenamefont {Geller}(1958)}]{Gilleo}%
  \BibitemOpen
  \bibfield  {author} {\bibinfo {author} {\bibfnamefont {M.~A.}\ \bibnamefont
  {Gilleo}}\ and\ \bibinfo {author} {\bibfnamefont {S.}~\bibnamefont
  {Geller}},\ }\href {\doibase 10.1103/PhysRev.110.73} {\bibfield  {journal}
  {\bibinfo  {journal} {Phys. Rev.}\ }\textbf {\bibinfo {volume} {110}},\
  \bibinfo {pages} {73} (\bibinfo {year} {1958})}\BibitemShut {NoStop}%
\bibitem [{\citenamefont {Xie}\ \emph {et~al.}(2017)\citenamefont {Xie},
  \citenamefont {Jin}, \citenamefont {He}, \citenamefont {Bauer}, \citenamefont
  {Barker},\ and\ \citenamefont {Xia}}]{Xie}%
  \BibitemOpen
  \bibfield  {author} {\bibinfo {author} {\bibfnamefont {L.-S.}\ \bibnamefont
  {Xie}}, \bibinfo {author} {\bibfnamefont {G.-X.}\ \bibnamefont {Jin}},
  \bibinfo {author} {\bibfnamefont {L.}~\bibnamefont {He}}, \bibinfo {author}
  {\bibfnamefont {G.~E.~W.}\ \bibnamefont {Bauer}}, \bibinfo {author}
  {\bibfnamefont {J.}~\bibnamefont {Barker}}, \ and\ \bibinfo {author}
  {\bibfnamefont {K.}~\bibnamefont {Xia}},\ }\href {\doibase
  10.1103/PhysRevB.95.014423} {\bibfield  {journal} {\bibinfo  {journal} {Phys.
  Rev. B}\ }\textbf {\bibinfo {volume} {95}},\ \bibinfo {pages} {014423}
  (\bibinfo {year} {2017})}\BibitemShut {NoStop}%
\bibitem [{\citenamefont {Prabhakar}\ and\ \citenamefont
  {Stancil}(2009)}]{spinwavebook}%
  \BibitemOpen
  \bibfield  {author} {\bibinfo {author} {\bibfnamefont {A.}~\bibnamefont
  {Prabhakar}}\ and\ \bibinfo {author} {\bibfnamefont {D.~D.}\ \bibnamefont
  {Stancil}},\ }\href@noop {} {\emph {\bibinfo {title} {Spin Waves: Theory and
  Applications}}}\ (\bibinfo  {publisher} {Springer, Boston, MA},\ \bibinfo
  {year} {2009})\BibitemShut {NoStop}%
\bibitem [{\citenamefont {{Fujisawa}}(1958)}]{Fujisawa}%
  \BibitemOpen
  \bibfield  {author} {\bibinfo {author} {\bibfnamefont {K.}~\bibnamefont
  {{Fujisawa}}},\ }\href {\doibase 10.1109/TMTT.1958.1125205} {\bibfield
  {journal} {\bibinfo  {journal} {IRE Transactions on Microwave Theory and
  Techniques}\ }\textbf {\bibinfo {volume} {6}},\ \bibinfo {pages} {344}
  (\bibinfo {year} {1958})}\BibitemShut {NoStop}%
\bibitem [{\citenamefont {Eshbach}(1962)}]{Eshbach}%
  \BibitemOpen
  \bibfield  {author} {\bibinfo {author} {\bibfnamefont {J.~R.}\ \bibnamefont
  {Eshbach}},\ }\href {\doibase 10.1103/PhysRevLett.8.357} {\bibfield
  {journal} {\bibinfo  {journal} {Phys. Rev. Lett.}\ }\textbf {\bibinfo
  {volume} {8}},\ \bibinfo {pages} {357} (\bibinfo {year} {1962})}\BibitemShut
  {NoStop}%
\bibitem [{\citenamefont {Le~Floch}\ \emph {et~al.}(2013)\citenamefont
  {Le~Floch}, \citenamefont {Fan}, \citenamefont {Aubourg}, \citenamefont
  {Cros}, \citenamefont {Carvalho}, \citenamefont {Shan}, \citenamefont
  {Bourhill}, \citenamefont {Ivanov}, \citenamefont {Humbert}, \citenamefont
  {Madrangeas},\ and\ \citenamefont {Tobar}}]{LeFloch}%
  \BibitemOpen
  \bibfield  {author} {\bibinfo {author} {\bibfnamefont {J.-M.}\ \bibnamefont
  {Le~Floch}}, \bibinfo {author} {\bibfnamefont {Y.}~\bibnamefont {Fan}},
  \bibinfo {author} {\bibfnamefont {M.}~\bibnamefont {Aubourg}}, \bibinfo
  {author} {\bibfnamefont {D.}~\bibnamefont {Cros}}, \bibinfo {author}
  {\bibfnamefont {N.~C.}\ \bibnamefont {Carvalho}}, \bibinfo {author}
  {\bibfnamefont {Q.}~\bibnamefont {Shan}}, \bibinfo {author} {\bibfnamefont
  {J.}~\bibnamefont {Bourhill}}, \bibinfo {author} {\bibfnamefont {E.~N.}\
  \bibnamefont {Ivanov}}, \bibinfo {author} {\bibfnamefont {G.}~\bibnamefont
  {Humbert}}, \bibinfo {author} {\bibfnamefont {V.}~\bibnamefont {Madrangeas}},
  \ and\ \bibinfo {author} {\bibfnamefont {M.~E.}\ \bibnamefont {Tobar}},\
  }\href {\doibase 10.1063/1.4848935} {\bibfield  {journal} {\bibinfo
  {journal} {Review of Scientific Instruments}\ }\textbf {\bibinfo {volume}
  {84}},\ \bibinfo {pages} {125114} (\bibinfo {year} {2013})}\BibitemShut
  {NoStop}%
\bibitem [{\citenamefont {Goryachev}\ \emph {et~al.}(2018)\citenamefont
  {Goryachev}, \citenamefont {Watt}, \citenamefont {Bourhill}, \citenamefont
  {Kostylev},\ and\ \citenamefont {Tobar}}]{goryachev18}%
  \BibitemOpen
  \bibfield  {author} {\bibinfo {author} {\bibfnamefont {M.}~\bibnamefont
  {Goryachev}}, \bibinfo {author} {\bibfnamefont {S.}~\bibnamefont {Watt}},
  \bibinfo {author} {\bibfnamefont {J.}~\bibnamefont {Bourhill}}, \bibinfo
  {author} {\bibfnamefont {M.}~\bibnamefont {Kostylev}}, \ and\ \bibinfo
  {author} {\bibfnamefont {M.~E.}\ \bibnamefont {Tobar}},\ }\href {\doibase
  10.1103/PhysRevB.97.155129} {\bibfield  {journal} {\bibinfo  {journal} {Phys.
  Rev. B}\ }\textbf {\bibinfo {volume} {97}},\ \bibinfo {pages} {155129}
  (\bibinfo {year} {2018})}\BibitemShut {NoStop}%
\bibitem [{\citenamefont {{Castel}}\ \emph {et~al.}(2019)\citenamefont
  {{Castel}}, \citenamefont {{Ammar}}, \citenamefont {{Manchec}}, \citenamefont
  {{Cochet}},\ and\ \citenamefont {{Youssef}}}]{castel19}%
  \BibitemOpen
  \bibfield  {author} {\bibinfo {author} {\bibfnamefont {V.}~\bibnamefont
  {{Castel}}}, \bibinfo {author} {\bibfnamefont {S.~B.}\ \bibnamefont
  {{Ammar}}}, \bibinfo {author} {\bibfnamefont {A.}~\bibnamefont {{Manchec}}},
  \bibinfo {author} {\bibfnamefont {G.}~\bibnamefont {{Cochet}}}, \ and\
  \bibinfo {author} {\bibfnamefont {J.~B.}\ \bibnamefont {{Youssef}}},\ }\href
  {\doibase 10.1109/LMAG.2019.2892927} {\bibfield  {journal} {\bibinfo
  {journal} {IEEE Magnetics Letters}\ }\textbf {\bibinfo {volume} {10}},\
  \bibinfo {pages} {1} (\bibinfo {year} {2019})}\BibitemShut {NoStop}%
\bibitem [{\citenamefont {Owens}(2019)}]{Clai}%
  \BibitemOpen
  \bibfield  {author} {\bibinfo {author} {\bibfnamefont {C.}~\bibnamefont
  {Owens}},\ }\emph {\bibinfo {title} {Creating Quantum Topological Materials
  with 3D Microwave Photons}},\ \href@noop {} {Ph.D. thesis},\ \bibinfo
  {school} {The University of Chicago} (\bibinfo {year} {2019})\BibitemShut
  {NoStop}%
\bibitem [{\citenamefont {Zhang}\ \emph {et~al.}(2017)\citenamefont {Zhang},
  \citenamefont {Luo}, \citenamefont {Wang}, \citenamefont {Li},\ and\
  \citenamefont {You}}]{NatCom8}%
  \BibitemOpen
  \bibfield  {author} {\bibinfo {author} {\bibfnamefont {D.}~\bibnamefont
  {Zhang}}, \bibinfo {author} {\bibfnamefont {X.-Q.}\ \bibnamefont {Luo}},
  \bibinfo {author} {\bibfnamefont {Y.-P.}\ \bibnamefont {Wang}}, \bibinfo
  {author} {\bibfnamefont {T.-F.}\ \bibnamefont {Li}}, \ and\ \bibinfo {author}
  {\bibfnamefont {J.~Q.}\ \bibnamefont {You}},\ }\href {\doibase
  10.1038/s41467-017-01634-w} {\bibfield  {journal} {\bibinfo  {journal}
  {Nature Communications}\ }\textbf {\bibinfo {volume} {8}},\ \bibinfo {pages}
  {1368} (\bibinfo {year} {2017})}\BibitemShut {NoStop}%
\bibitem [{\citenamefont {Harder}\ \emph {et~al.}(2018)\citenamefont {Harder},
  \citenamefont {Yang}, \citenamefont {Yao}, \citenamefont {Yu}, \citenamefont
  {Rao}, \citenamefont {Gui}, \citenamefont {Stamps},\ and\ \citenamefont
  {Hu}}]{Harder}%
  \BibitemOpen
  \bibfield  {author} {\bibinfo {author} {\bibfnamefont {M.}~\bibnamefont
  {Harder}}, \bibinfo {author} {\bibfnamefont {Y.}~\bibnamefont {Yang}},
  \bibinfo {author} {\bibfnamefont {B.~M.}\ \bibnamefont {Yao}}, \bibinfo
  {author} {\bibfnamefont {C.~H.}\ \bibnamefont {Yu}}, \bibinfo {author}
  {\bibfnamefont {J.~W.}\ \bibnamefont {Rao}}, \bibinfo {author} {\bibfnamefont
  {Y.~S.}\ \bibnamefont {Gui}}, \bibinfo {author} {\bibfnamefont {R.~L.}\
  \bibnamefont {Stamps}}, \ and\ \bibinfo {author} {\bibfnamefont {C.-M.}\
  \bibnamefont {Hu}},\ }\href {\doibase 10.1103/PhysRevLett.121.137203}
  {\bibfield  {journal} {\bibinfo  {journal} {Phys. Rev. Lett.}\ }\textbf
  {\bibinfo {volume} {121}},\ \bibinfo {pages} {137203} (\bibinfo {year}
  {2018})}\BibitemShut {NoStop}%
\bibitem [{\citenamefont {Match}\ \emph {et~al.}(2019)\citenamefont {Match},
  \citenamefont {Harder}, \citenamefont {Bai}, \citenamefont {Hyde},\ and\
  \citenamefont {Hu}}]{PhysRevB.99.134445}%
  \BibitemOpen
  \bibfield  {author} {\bibinfo {author} {\bibfnamefont {C.}~\bibnamefont
  {Match}}, \bibinfo {author} {\bibfnamefont {M.}~\bibnamefont {Harder}},
  \bibinfo {author} {\bibfnamefont {L.}~\bibnamefont {Bai}}, \bibinfo {author}
  {\bibfnamefont {P.}~\bibnamefont {Hyde}}, \ and\ \bibinfo {author}
  {\bibfnamefont {C.-M.}\ \bibnamefont {Hu}},\ }\href {\doibase
  10.1103/PhysRevB.99.134445} {\bibfield  {journal} {\bibinfo  {journal} {Phys.
  Rev. B}\ }\textbf {\bibinfo {volume} {99}},\ \bibinfo {pages} {134445}
  (\bibinfo {year} {2019})}\BibitemShut {NoStop}%
\bibitem [{\citenamefont {Goryachev}\ \emph {et~al.}(2017)\citenamefont
  {Goryachev}, \citenamefont {Kostylev},\ and\ \citenamefont
  {Tobar}}]{goryachev2017strong}%
  \BibitemOpen
  \bibfield  {author} {\bibinfo {author} {\bibfnamefont {M.}~\bibnamefont
  {Goryachev}}, \bibinfo {author} {\bibfnamefont {M.}~\bibnamefont {Kostylev}},
  \ and\ \bibinfo {author} {\bibfnamefont {M.~E.}\ \bibnamefont {Tobar}},\
  }\href@noop {} {\enquote {\bibinfo {title} {Strong coupling of 3d cavity
  photons to travelling magnons at low temperatures},}\ } (\bibinfo {year}
  {2017}),\ \Eprint {http://arxiv.org/abs/1710.06601} {arXiv:1710.06601
  [cond-mat.mtrl-sci]} \BibitemShut {NoStop}%
\end{thebibliography}
%

\newpage
\clearpage
\beginsupplement
\section*{Supplementary Material}
\subsection{Cavity $Q$ factors}
	
\begin{figure}[h!]
		\centering
		\includegraphics[width=0.35\textwidth]{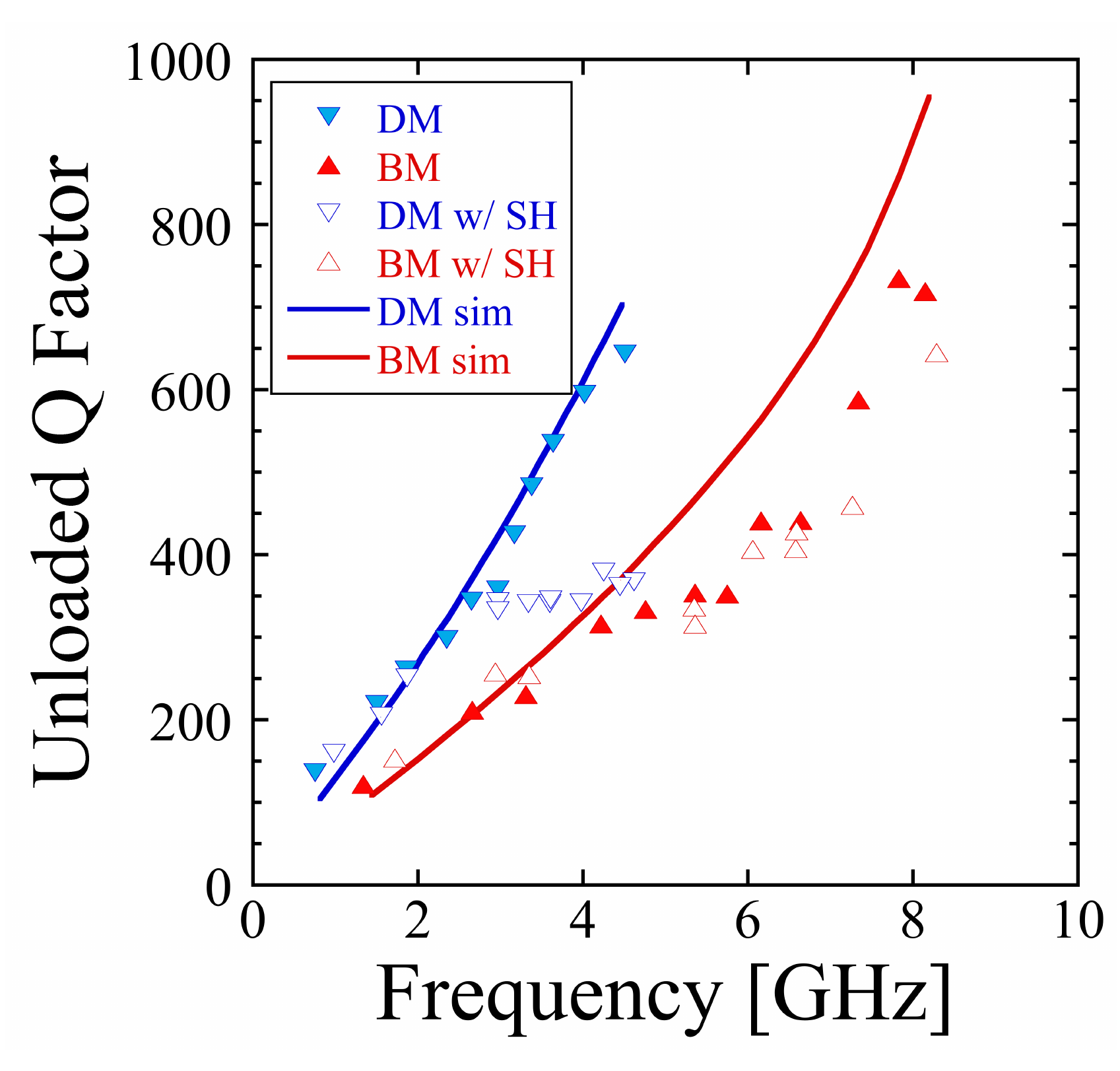}
		\caption{Unloaded $Q$ factor as a function of mode frequency for the DM and BM, with and without the sample holder (SH) in the cavity. Simulated values are obtained from calculating the GF of the cavity and then scaling by $1/R_s$ to fit the data.}
		\label{fig:Q}
	\end{figure}

An important parameter of any microwave cavity is the so-called Geometric Factor, which can be calculated as
	\begin{equation}
		 \text{GF}=\mu_0\omega_c\frac{\int_{V_c}|H|^2dV}{\int_{S_c}|H|^2dS}.
	\end{equation}
This can be used to estimate cavity $Q$ factors as $Q=\text{GF}/R_s$, where $R_s$ is the surface resistance of the cavity walls. It represents the ratio of the magnetic field within the volume of the cavity over the field on the metallic surface, where there exists some small resistance and hence microwave loss. Measured cavity $Q$ factors are plotted in Fig.~\ref{fig:Q}, with simulated GF factors shown as well, which have been scaled to fit by $R_s=0.09~\Omega$. We observe that the BM is unaffected by the presence of the sample holder in the cavity, whereas it becomes the dominant loss mechanism for the DM for $Q$ factors above $\sim350$, which is due to a higher electrical filling in the sample holder for the DM hence a higher impact of its dielectric loss.

\subsection{Measuring magnon linewidth}

\begin{figure}[h!]
	\centering
	\includegraphics[width=0.43\textwidth]{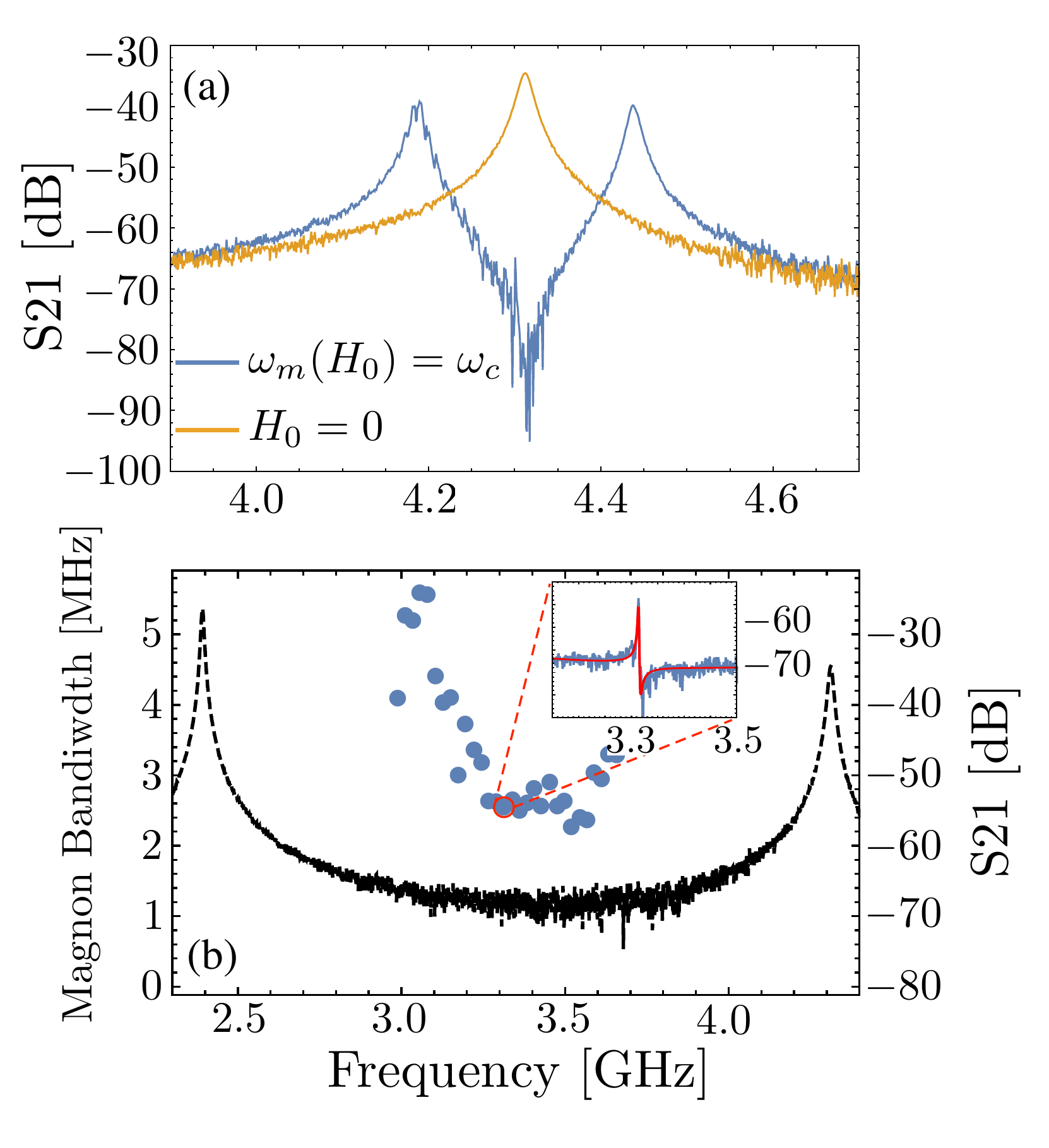}
	\caption{(a) Transmission spectra for the BM--CP magnon and cavity tuned (blue) and detuned (yellow) demonstrated the characteristic mode splitting caused by strong coupling. (b) Measurements of the BM--CP magnon linewidth at various magnetic field values such that the magnon frequency sits between the two cavity modes.}
	\label{fig:linewidth}
\end{figure}

Aside from the characteristic mode splitting demonstrated in Fig.~\ref{fig:linewidth}(a), to fulfil the condition of strong coupling, a cavity--magnon system must have its coupling value $g_{cm}$ larger than both the cavity and magnon linewidths. We can approximate cavity linewidths from the inverse of the gradients in Fig.~\ref{fig:Q} as $\kappa_{DM}=6.8$ MHz and $\kappa_{BM}=10.5$ MHz. 

Magnon linewidth can be measured when the magnetic resonance is tuned approximately halfway between the two cavity modes. As the magnon frequency approaches the cavity modes it begins to hybridise with them and hence its linewidth becomes closer to that of the cavity mode{'}s, as seen in Fig.~\ref{fig:linewidth}(b). However, at its minimum value, a Fano resonance fit is used to determine that $\kappa_{m}\approx2.5$ MHz for the thin film sample, as seen in the inset image in Fig.~\ref{fig:linewidth}(b). 

Given that our smallest measured coupling values were $g_{cm}\approx20$ MHz, the system therefore fulfils the condition for strong coupling under all circumstances tested here. 

\subsection{Calculating $\eta$}
The various integrals in the formula for the factor $\eta$ can be calculated from finite element modelling. Their values are given in the table below. An {``}ultra-fine{''} mesh was used in COMSOL$^\text{TM}$, and the dielectric constant of YIG included for the magnetic sample volume. 

\begin{table*}[!htb]

\smallskip 
\begin{tabular*}{\textwidth}{@{\extracolsep{\fill}}clllllll}
\toprule
Case & $d1~(=d2)$ & $\omega_c/2\pi$ & $\int \vec{H} . \vec{e_x} dV_m$ & $\int \vec{H} . \vec{e_y} dV_m$ & $\int |\vec{H}|^2dV_c$ & $V_m$ & $\eta$ \\
& [mm] & [GHz] & [m$^2$A $\times10^{-9}$] & [m$^2$A $\times10^{-9}$] & [mA$^2$ $\times 10^{-4}$]  & [$\times 10^{-10}$ m$^3$] &[$\times10^{-2}$]\\
\hline
DM-OCP& 0.040 & 1.637 & 2.24 & 5.73 & 1.98 & 4.61 & 2.02\\
DM-OCP& 0.064 & 1.898 & 1.93 & 4.89 &1.48 & 4.61 & 2.00\\
DM-OCP & 0.100 & 2.264 & 1.58 & 4.03 & 1.05 & 4.61 & 1.96 \\
DM-OCP & 0.126 & 2.458 & 1.56 & 3.97 &1.05 & 4.61 & 1.93 \\
DM-OCP & 0.398 & 3.530 & 2.18 & 5.53 & 2.61 & 4.61 & 1.70 \\
DM-OCP & 1.500 & 4.754 & 1.76 & 4.35 & 3.20 & 4.61 & 1.21 \\
\hline
BM-OCP & 0.0400 & 2.769 & 1.92 & 3.61 &2.16 & 4.61 & 1.29\\
BM-OCP & 0.100 & 4.070 & 1.28 & 2.23 & 1.01 & 4.61 & 1.18\\
BM-OCP & 0.1250 & 4.441 & 1.26 & 2.13 & 1.01 & 4.61 & 1.14\\ 
BM-OCP & 0.300 & 5.993 & 1.71 & 2.23 &2.20 & 4.61 &0.876\\
BM-OCP & 0.500 & 6.911 & 1.76 & 1.52 & 2.96 & 4.61 & 0.627\\ 
BM-OCP & 0.900 & 7.849& 1.39 & 1.56 &2.96 & 4.61 & 0.561 \\
BM-OCP & 1.500 & 8.421 & 1.17 & 2.37 & 3.51 & 4.61 & 0.651 \\
\hline
BM-CP & 0.0199 & 2.024 & 0.292  & 6.64  & 0.512 & 4.61 & 4.33\\
BM-CP & 0.0316 & 2.502 & 0.438  & 6.00  & 0.338 & 4.61 & 4.33\\
BM-CP & 0.0502 & 3.069 & 0.360  & 4.45  & 0.230 & 4.61 & 4.34\\
BM-CP & 0.0794 & 3.727 & 0.355  & 4.39  & 0.224 & 4.61 & 4.34\\
BM-CP & 0.126 & 4.472 & 0.368  & 4.61  & 0.247 & 4.61 & 4.33\\
BM-CP & 0.199 & 5.290 & 0.378  & 4.75  & 0.264 & 4.61 & 4.32\\
BM-CP & 0.316 & 6.152 & 0.467  & 6.03  & 0.439 & 4.61 & 4.26\\
BM-CP & 0.501 & 7.019 & 0.493  & 6.56  & 0.555 & 4.61 & 4.11\\
BM-CP & 0.794 & 7.838 & 0.467  & 6.56  & 0.655 & 4.61 & 3.78\\

\hline
BM-CP [SPHERE] & 0.0126 & 1.63 & 2.80 $\times10^{-3}$ & 1.46& 0.866 & 0.544 & 2.12\\
BM-CP [SPHERE] & 0.0158 & 1.82 & 2.51 $\times10^{-3}$ & 1.30 & 0.693 & 0.544 & 2.12\\
BM-CP [SPHERE] & 0.0199 & 2.02 & 2.25 $\times10^{-3}$ & 1.17 &0.557 & 0.544 & 2.12\\
BM-CP [SPHERE] & 0.0251 & 2.25 & 2.02 $\times10^{-3}$ & 1.05 & 0.451 & 0.544 & 2.12 \\
BM-CP [SPHERE] & 0.0794 & 3.73 & 1.51 $\times10^{-3}$ & 0.775 & 0.249 & 0.544 & 2.10\\
BM-CP [SPHERE] & 0.1 & 4.09 & 1.65 $\times10^{-3}$ & 0.842 & 0.295 & 0.544 & 2.10\\
BM-CP [SPHERE] & 0.158 & 4.88 & 1.69 $\times10^{-3}$ & 0.859 & 0.313 & 0.544 & 2.08\\
BM-CP [SPHERE] & 0.199 & 5.30 & 1.89 $\times10^{-3}$ & 0.957 & 0.394 & 0.544 & 2.06\\
\hline
\label{tab:eta}
\end{tabular*}
\caption{FEM derived values for magnetic field integrals for each of the four spectroscopy test cases and the resulting value of $\eta$.}
\end{table*}

%
%

\end{document}